\documentclass[11pt]{article}
\usepackage{amsfonts}
\usepackage{latexsym}

\usepackage{subfigure}
\usepackage{graphics}

\usepackage[dvips,final]{epsfig}
\usepackage{amsfonts}
\usepackage{latexsym}

\usepackage{subfigure}
\usepackage{color}

\newcommand{\qa}{\mbox{\quad\mbox{and}\quad}}

 \newcommand{\C}{\mathbb{C}}
  \newcommand{\cR}{\mathcal{R}}
  \newcommand{\rank}{\mathrm{rank\;}}

\newcommand{\g}{\mbox{$\cal G$}}

\newcommand{\kk}{\mbox{$\cal K$}}

\newcommand{\oo}{\mbox{$\mathbb O$}}

\newcommand{\x}{\mbox{$\mathbf x$}}
\newcommand{\y}{\mbox{$\mathbf y$}}

\newcommand{\zz}{\mbox{$\mathbf z$}}

\newcommand{\rt}{\mbox{$\mathbb R$}}

\newcommand{\ttt}{\mbox{$\cal T$}}
\newcommand{\f}{\mbox{$\cal F$}}

 \newcommand{\diag}{\mathrm{diag}}

\newcommand{\aab}{\mbox{$\mathbf a$}}

\newcommand{\dr}{\mbox{\tiny $\Delta$}}
\newcommand{\er}{\mbox{$\varepsilon$}}

\newtheorem{remark}{Remark}
\newtheorem{theorem}{Theorem}

\newtheorem{corollary}{Corollary}

\newtheorem{proof}{Proof}

\date{}

\begin{document}

 \title{\bf  A new  technique for compression of data sets }

  \author{Anatoli Torokhti\thanks{Centre for Industrial and Applied
  Mathematics, School of Mathematics and Statistics,  University of South  Australia, SA 5095, Australia. Ph: +61 8 8302 3812.
 Fax: +61 8 8302 5785.  E-mail: anatoli. torokhti@unisa.edu.au.} }

 \maketitle
\begin{abstract}
Data compression techniques are characterized by  four key performance indices  which are (i) associated accuracy, (ii) compression
ratio, (iii) computational work, and (iv) degree of freedom.
The method of data compression developed in this paper allows us to substantially improve all the four issues
 above.

The proposed transform $\f$ is presented in the form of a sum with $p-1$ terms, $\f_1,\ldots, \f_{p-1}$, where each term
is a particular sub-transform presented by a first degree polynomial. For $j=1,\ldots,p-1$, each sub-transform $\f_{j}$ is determined from
interpolation-like conditions. This device provides the transform  flexibility  to incorporate variation of observed data and leads
to  performance improvement. The transform $\f$ has two degrees of freedom, the number of sub-transforms and associated
compression ratio associated with each sub-transform $\f_{j}$.
\end{abstract}


\section{Introduction}\label{int}


{I}n this paper, a new  technique for compression  of a  set of random signals is proposed. The technique
provides the better associated accuracy, compression ratio and computational load than  the Karhunen-Lo\`{e}ve
transform (KLT) and its known extensions.

The basic idea is to {\em  combine}  the methodologies  of a piece-wise liner
function interpolation \cite{mei1} and the best rank-constrained operator approximation.  This issue is discussed in more detail
in Sections \ref{gentle} and \ref{statement}.

{T}{he} KLT  is the fundamental and, perhaps, most popular data compression method \cite{sch1}--\cite{tor-m2}. It is also
known as the Principal Component Analysis (PCA) \cite{jol1,tor-f1} and ``is probably the oldest and best known of the techniques
of multivariate analysis" \cite{jol1}.
This technique   is intensively used
in a vide range of
research areas such as data compression \cite{jon13,sim14}, pattern recognition \cite{lun11}, interference suppression
\cite{hon2}, image processing \cite{bas11}, 
forecasting \cite{sto1},
 hydrology \cite{zha11}, physics nonlinear phenomena \cite{sch11},
probabilistic mechanics \cite{pho12,gri11}, 
biomechanics \cite{rap11},
geoscience \cite{jie11,pen11}, 
stochastic processes \cite{deh12,gas15}, information theory \cite{gas11,eff11}, 3-D object classification and video coding \cite{gao11},
chemical theory \cite{hou13}, oceanology \cite{che14}, optics and laser technology \cite{dub22}  and others.

The purpose of techniques based on the KLT is to compress observable data vector $\y$ with $n$ components to a shorter vector
$\widetilde{\x}$ with $r$ components (also called the principal components \cite{jol1}) and then reconstruct it so that the
reconstruction is close to the reference signal $\x$.
In general, $\x$ contains $m$ components where $m$ is not necessarily equal to $n$. The ratio
$$\displaystyle c=\frac{r}{m},\quad \mbox{where $r\leq m$,}
$$
is called the compression ratio. Smaller $r$ implies poorer accuracy in the reconstruction of the compressed data. In this sense,
$r$ is the KLT degree of freedom.
Therefore, the performance of the KLT and related  transforms is characterized by  four key performance indices which are
(i)  { associated accuracy},
 (ii) { compression ratio, }
 (iii) {  computational work} and (iv) the degree of freedom.

The method of data compression developed in this paper allows us to substantially (i) increase  the accuracy, (ii) improve
the compression ratio, (iii) decrease computational work and (iv) increase the degrees of freedom.

\subsection{Motivations for the proposed technique}\label{mot}

The  motivations of the proposed are as follows.
\subsubsection{Infinite sets of signals}\label{} Most of the literature on the subject of the KLT\footnote{Relevant
references can be found, for example, in \cite{tor100}--\cite{tor-m2}.}
discusses the properties of an optimal transform for an {\em individual} {\em finite} random signal.\footnote{Here, the signal is
treated as a vector with random components. We say
that a random signal $\x$ is finite if $\x$ has a finite number of components. } This means that if one wishes to compress and then
reconstruct an {\em infinite}  set of observable signals $K_{_Y}=\{\y_1, \y_2, \ldots, \y_N, \ldots\}$
{ so that reconstructed signals are close to an {\em infinite}  set of reference random signals $K_{_X}=\{\x_1,
\x_2, \ldots, \x_N, \ldots\}$ }
using the KLT approach then one is forced to find an {\em infinite} set of corresponding transforms
 $\{\ttt_1, \ttt_2, \ldots,\ttt_N, \ldots\}$: one element $\ttt_i$ of the transform set for each representative $\y_i$ of the
signal set $K_{_Y}$.
Clearly, such an approach cannot be
applied in practice.\footnote{Here, $K_{_Y}$ and $K_{_X}$ are countable sets. More generally, the sets $K_{_Y}$ and
$K_{_X}$ might be uncountable  when $K_{_Y}$ and $K_{_X}$ depend on a continuous parameter $\alpha$.}
 Although the standard KLT has been extended in \cite{tor1} to the cases of infinite signal sets, its associated
 accuracy is still not satisfactory (see Sections \ref{acc1}, \ref{comp} and \ref{sim}).

\subsubsection{Computational work}\label{} { Note that even in the case when
$K_{_Y}=\{\y_1, \y_2, \ldots, $ $\y_N\}$ and $K_{_X}=\{\x_1, \x_2, \ldots, \x_N\}$ with $N$ fixed
can be represented as finite `long' signals,
the KLT  applied to such signals leads to computation of large covariance matrices.  Indeed, if each $\y_i$ has $n$ components
and each $\x_i$ has $m$ components then the KLT approach leads to computation of a product of an $mN \times
nN$ matrix and an $nN \times nN$ matrix and computation of a $nN \times nN$ pseudo-inverse matrix \cite{tor100}.
This requires $O(2mn^2N^3)$ and $O(22n^3N^3)$ flops, respectively \cite{gol1}. As a result, in this case, the computational
work associated with the KLT approach becomes unreasonably hard.}

\subsubsection{Associated accuracy}\label{acc1} For the given compression ratio, the accuracy associated with the KLT-like
techniques cannot be changed. If the accuracy is not satisfactory then one has to change the compression ratio, which can be
undesirable. Thus, for the KLT approach, the compression ratio is the only degree of freedom to improve the accuracy.

\subsection{Brief description of the KLT }\label{}

First, we need some notation as follows.
Let $X=(\Omega, \Sigma, \mu)$ be a probability space, where $\Omega = \{\omega\}$ is the set of outcomes,
$\Sigma$ a $\sigma$--field of measurable subsets in $\Omega$ and $\mu:\Sigma \rightarrow [0,1]$ an associated
probability measure on $\Sigma$.

Let $\x \in L^{2}(\Omega,{\mathbb R}^{n})$ and $\y\in L^{2}(\Omega,{\mathbb R}^{m})$  be a reference signal and
an observed signal, respectively. Let the norm $\|\cdot\|^2_{\Omega}$ be given by
$
\displaystyle \|\x\|^2_{\Omega} = {\int}_{\Omega} \|\x(\omega)\|^2_2,
d\mu (\omega)
$
where $\|\x(\omega)\|_2$ is the Euclidean norm of $\x(\omega) \in {\mathbb R}^{m}$.

We consider a linear operator (transform) $\kk:$ $L^{2}(\Omega,{\mathbb R}^{m})$ $\rightarrow L^{2}(\Omega,{\mathbb R}^{n})$ defined by
a matrix $K\in \rt^{m\times n}$ so that
\begin{equation} \label{kk}
[\kk(\y)](\omega) = K[\y(\omega)].
\end{equation}
A generic KLT \cite{tor100} is the linear transform $\kk$ that solves
\begin{equation} \label{mmm}
\min_{\mathbb K} \|\x - \kk (\y)\|^2_\Omega \quad \mbox{subject to $\rank \kk = r\leq \min \{m, n\}$},
\end{equation}
where $\mathbb K$ is a class of all linear operators defined by (\ref{kk}).
As a result, the KLT returns two matrices, $D\in\rt^{m\times r}$ and $C\in\rt^{r\times n}$,
so that $K=DC$. Matrix  $C$ performs filtering and compression of observed data $\y$ to a shorter vector
$\widetilde{\x}\in L^{2}(\Omega,{\mathbb R}^{r})$  with $r$ components.
Matrix $D$ reconstructs $\widetilde{\x}$ in such a way that the reconstructed signal is close to $\x$ in the sense of
(\ref{mmm}).

Operator $\kk$ that solves  (\ref{mmm}) for a particular case  with $\y=\x$ is the standard KLT.

Thus, for a given compression ratio,  the KLT  minimizes the associated error over the class of all {\em linear} transforms of
the rank defined by (\ref{mmm}).

\subsection{Related techniques }\label{}

Despite the known KLT optimality, it may happen that the accuracy  and compression ratio associated with the KLT are
still not satisfactory. Owing to this observation and  the KLT versatility in applications  (see
Section \ref{int} above), the KLT idea was
extended in different directions as  has been done, in particular, in \cite{sch1}--\cite{hua2}. More recent
developments in this area include the polynomial transforms \cite{tor100},  extensions of the KLT  to the cases of infinite signal
sets \cite{tor1}, the best weighted  estimators
\cite{tor-m1,tor-f1}, distributed KLT \cite{gas11} and the   fast KLT using  wavelets
\cite{can11}.

Nevertheless, the known transforms based on the KLT idea still imply intrinsic difficulties associated
  with the original KLT. In particular, most of them have  been developed for the case of a {\em single} signal, and
for the fixed compression ratio, the associated error  cannot be improved.
In the case of the polynomial transforms \cite{tor100}, the error can be improved if the number of terms in the polynomial
transforms increase. The latter implies a substantial  computational
 burden that, in many cases,  may stifle this intention.

\subsection{Contribution. Particular features of the proposed transform}\label{con}

To describe the contribution and compare the features of the proposed transform  with those of the KLT-like techniques, we
consider the same issues as in Section \ref{mot}.
\subsubsection{Infinite sets of signals}\label{} Unlike most of the known transforms, the proposed technique allows us
to determine   {\em a single} transform to compress any signal from the {\em infinite} signal set. We note that {\em infinite} sets of
stochastic signal-vectors introduced in Section \ref{some}  are quite large and include, for example, time series.

\subsubsection{Computational work}\label{} In the case of finite signal sets, the proposed technique requires a lesser  computational load compared to
the KLT approach (see Sections \ref{numer} and \ref{sim}).

\subsubsection{Associated accuracy}\label{} Our transform provides data compression and  subsequent reconstruction with any
desired accuracy (Theorem \ref{thm3} in Section \ref{error})\footnote{This means that any desired accuracy is achieved theoretically,
 as is shown in Section \ref{error}
below. In practice, of course, the accuracy is increased to a prescribed reasonable level.}. This is achieved due to
the transform structure that follows from the device of the piece-wise function interpolation (Sections \ref{some}
and \ref{statement}). In particular,
it provides the related degree of freedom, the number of the so-called interpolation pairs (see Sections \ref{statement} and
\ref{degree}), to improve the transform performance (Section  \ref{comp}).

Moreover, the proposed  technique

(i) determines the data compression transform in terms of pseudo-inverse matrices so that the transform always exists,

\hspace*{-2mm}  (ii)  uses the same  initial information (signal samples) as is used in the KLT-like transforms (see
Sections \ref{part11} and \ref{sim}), and

\hspace*{-2mm}  (iii) provides the simultaneous filtering and compression.

The paper is organized as follows. In Section \ref{gentle}, a brief description of the problem is given.
 Some necessary preliminaries, used in our transform construction,   are given in
Section \ref{some}. In Section \ref{filter},  the transform model $\f$ is provided. In Section
\ref{statement}, a rigorous statement of the problem is given, which is a generalization and extension of problem
(\ref{mmm}) to the case of filtering of infinite sets of stochastic signals.  In Section \ref{main},
the proposed transform  $F$ is determined from interpolation conditions, and the associated error analysis is presented.
In particular, in Section \ref{comp}, a comparison with KLT-like approach is given.

\section{Underlying  idea. Device of the transform}\label{gentle}

\subsection{Underlying  idea}

In this paper, the  underlying idea is different from those considered in the above mentioned works. The proposed transform
is constructed from {\em a combination} of the device of the piece-wise linear
function interpolation \cite{mei1} and the best rank-constrained operator approximation. This means the following.

Suppose, $K_{_Y}$ and $K_{_X}$ are uncountable sets of random signals. While the rigorous notation is given in Section
\ref{some}, here, we denote $\y(t,\cdot)\in K_{_Y}$ and $\x(t,\cdot)\in K_{_X}$ where $t\in [a\hspace*{2mm}  b]\subset \rt$
is a time snapshot. Let $a= t_1 \leq \ldots \leq t_p = b.$
Choose  finite numbers of signals,  $\{\y(t_1,\cdot),\ldots, \y(t_p,\cdot)\}\subset K_{_Y}$ and
$\{\x(t_1,\cdot),\ldots, \x(t_p,\cdot)\}\subset K_{_X}$. The proposed transform $\f$ contains $p-1$ terms $\f_1,\ldots,
\f_{p-1}$ (see (\ref{fyd})--(\ref{fydj}) below) where, for $j=1,\ldots,p-1$, each term $\f_j$ is given as a first order
operator polynomial (see (\ref{fydj}) below) and is determined from the interpolation
conditions
\begin{equation} \label{cond1}
\hspace*{-2mm}\f_j [\y(t_j,\cdot)] = \x(t_j,\cdot) \hspace*{2mm}\mbox{and}\hspace*{2mm} \f_j [\y(t_{j+1},\cdot)] \approx \x(t_{j+1},\cdot).
\end{equation}

A reason for using the approximate equality in (\ref{cond1}) is explained in Section \ref{diss}. In Section \ref{prel}, the second
condition in  (\ref{cond1}) is represented as the best rank-constrained approximation problem
 (\ref{min1})--(\ref{r1}). As a result, such a transform has advantages  that are similar to the known advantages  of the piece-wise
 function interpolation as  has been mentioned in Section \ref{con}.
The procedure of data compression-reconstruction is performed via $p-1$ truncated singular
value decompositions (SVD) provided in Section \ref{part}.

\subsection{Device of the transform}\label{some}
\subsubsection{Signal sets under consideration}

We wish to consider signals in some wide sense as follows. Let
$T:=[a,\hspace*{1mm} b]\subseteq \rt$, and let $K_{_X}$ and $K_{_Y}$  be infinite sets of signals,\\
 $K_{_X}=\{ \x(t, \cdot) \in
L^{2}(\Omega,{\mathbb R}^{m})\hspace*{1mm}| \hspace*{1mm} t \in
T\}$ and $K_{_Y}=\{ \y(t, \cdot)$ $ \in L^{2}(\Omega,{\mathbb
R}^{n})\hspace*{1mm}| \hspace*{1mm} t \in T\}$ where
 \begin{eqnarray*}
&&\x(t, \cdot) = [\x^{(1)}(t, \cdot) \ldots \x^{(m)}(t, \cdot)]^T \\
&&\hspace*{-10mm}\mbox{with}\quad \x^{(j)}(t, \cdot)
 \in L^{2}(\Omega,{\mathbb R})\quad
\forall \quad j=1,\ldots,m,
\end{eqnarray*}
and
 \begin{eqnarray*}
&& \y(t, \cdot) = [\y^{(1)}(t, \cdot) \ldots \y^{(n)}(t, \cdot)]^T \\
&&\hspace*{-10mm} \mbox{with}\quad \y^{(i)}(t, \cdot)
  \in L^{2}(\Omega,{\mathbb R})
 \quad \forall \quad j=1,\ldots,n,
\end{eqnarray*}
respectively.

We interpret $\x(t, \cdot)$ as a reference signal  and $\y(t, \cdot)$ as an observable signal.\footnote{In an intuitive way $\y(t, \cdot)$ can be
 regarded as a noise-corrupted version of $\x(t, \cdot)$. For example, $\y(t, \cdot)$ can be interpreted as $\y(t, \cdot) =
 \x(t, \cdot) +{\mathbf n}$ where ${\mathbf n}$ is  white noise. In this paper, we do not restrict ourselves  to this simplest version of
 $\y(t, \cdot)$ and assume that the dependence of $\y(t, \cdot)$ on $\x(t, \cdot)$ and  ${\mathbf n}$ is arbitrary.}

The variable $t \in T$ represents time.\footnote{More generally, $T$ can be
considered as a set of  parameter vectors $\alpha = (\alpha^{(1)}, \ldots, \alpha^{(q)})^T \in C^q\subseteq
\rt^q$, where $C^q$ is a $q$-dimensional cube. One coordinate, say $\alpha^{(1)}$ of $\alpha$, could be interpreted as time.}
Then, for example, $\x(t, \cdot)$  can be interpreted as an arbitrary stationary time series.


Let $\{t_k\}_1^p\subset T$ be a nondecreasing sequence of fixed time-points such that
\begin{equation}\label{atjb}
a = t_1 \leq \ldots \leq t_p = b.
\end{equation}

\subsubsection{The transform model}\label{filter}

Now let $\f:L^2(\Omega,\rt^n) \rightarrow L^2(\Omega,\rt^m)$ be a transform such that, for each $t\in T$,
\begin{equation}\label{fyd}
\f[\y(t,\cdot)] = \sum_{j=1}^{p-1}\delta_j  \f_j [\y(t,\cdot)],
\end{equation}
 where
 \begin{equation}\label{fydj}
 \f_j[\y(t,\cdot)] = \aab_j + \g_j[\y(t,\cdot)]\hspace*{2mm}\mbox{and}\hspace*{2mm}  \delta_j = \left\{ \begin{array}{cl}
                             1,  & \hspace*{-2mm}\mbox{if $t_j\leq t \leq t_{j+1}$},\\
                             0, & \hspace*{-2mm}\mbox{otherwise.}
                              \end{array} \right.
\end{equation}
Here, $\f_j$  is a sub-transform with $\aab_j\in  L^2(\Omega,\rt^m) $ and $\g_j: L^2(\Omega,\rt^n) \rightarrow L^2(\Omega,\rt^m)$
is a linear operator represented by a matrix $G_j\in\rt^{m\times n}$ so that
$$
[\g_j(\y)](t,\omega) = G_j[\y(t,\omega)].
$$
 For each $j=1,\ldots,p-1$, vector $\aab_j$ and operator $\g_j$ should be determined from the interpolation conditions given in
 the following Section \ref{statement}.

Thus, $\f_j$ is defined by a matrix $F_j\in\rt^{m\times n}$ such that
\begin{equation}\label{fyab}
F_j[\y(t,\omega)] = \aab_j(\omega) + G_j[\y(t,\omega)].
\end{equation}


\section{Statement of the problem}\label{statement}

Let $\x(t_{1},\cdot),\ldots, \x(t_{p},\cdot)$ and $\y(t_{1},\cdot),\ldots, \y(t_{p},\cdot)$ be signals chosen
from infinite signal sets $K_{_X}$ and $K_{_Y}$.

\subsection{Preliminaries}\label{prel}
 Ideally, in (\ref{fyd})--(\ref{fyab}), for $j=1,\ldots,p-1$, we would like to determine each $\f_j$ so that $\aab_j$ satisfies
\begin{eqnarray}\label{f1j}
\aab_j + \g_j[\y(t_j,\cdot)] = \x(t_j,\cdot)
\end{eqnarray}
and $\g_j$ solves
\begin{eqnarray}\label{f2j}
 \aab_j + \g_j[\y(t_{j+1},\cdot)]= \x(t_{j+1},\cdot).
\end{eqnarray}
The conditions (\ref{f1j})--(\ref{f2j}) are motivated by the device of piece-wise function interpolation and associated advantages
\cite{mei1}. In turn, (\ref{f1j}) implies $\aab_j  = \x(t_j,\cdot) - \g_j [\y(t_j,\cdot)]$ which being substituted in (\ref{f2j}),
reduces (\ref{f2j}) to
\begin{eqnarray}\label{drj}
&& \dr\x(t_j,t_{j+1},\cdot) - \g_j [\dr\y(t_j,t_{j+1},\cdot)] = \theta,
\end{eqnarray}
where \begin{eqnarray*}
&&\dr\x(t_j,t_{j+1},\cdot)= \x(t_{j+1},\cdot)-\x(t_{j},\cdot), \\
&&\dr\y(t_j,t_{j+1},\cdot)= \y(t_{j+1},\cdot)-\y(t_{j},\cdot)
\end{eqnarray*}
 and $\theta$ is the zero vector.

Nevertheless, the transform $\f$ with such $\aab_j$ and  $\g_j$ does not provide data compression. To provide compression, we
  require that instead of (\ref{drj}), $\g_j$ solves
\begin{eqnarray}\label{min1}
\min_{\small\g_j} \left \|\dr\x(t_j,t_{j+1},\cdot) - \g_j [\dr\y(t_j,t_{j+1},\cdot)]\right \|^2_\Omega
\end{eqnarray}
subject to
\begin{eqnarray}\label{r1}
\rank \g_j  = r_j\leq \min \{m, n\}.
\end{eqnarray}
In (\ref{min1}), we use the notation
\begin{equation} \label{nx}
\|\dr\x(t_j,t_{j+1},\cdot)\|^2_{\Omega} = {\int}_{\Omega} \|\dr\x(t_j,t_{j+1},\omega)\|^2_2 d\mu (\omega).
\end{equation}
The constraint (\ref{r1}) leads to compression of $\y(t,\cdot)$  to a shorter vector
with $r_j$ components. This issue is discussed in more detail in Section \ref{part} below.

\subsection{Problem formulation}

Thus,  a determination of  $\f$ presented by (\ref{fyd})--(\ref{fyab}) is reduced to {\em  the following problem:}
Given $\{\x(t_j,\cdot), \y(t_j,\cdot)\}_{j=1}^{p-1}$, let
  for $j=1,\ldots,p-1$,
\begin{eqnarray}\label{axj}
\aab_j  = \x(t_j,\cdot) - \g_j [\y(t_j,\cdot)],
\end{eqnarray}
as in (\ref{f1j}). Find $\g_j$ that solves (\ref{min1}) subject to (\ref{r1}).

The above term ``given" means that covariance matrices associated with $\x(t_j,\cdot)$ are  $\y(t_j,\cdot)$ are known
or can be estimated. This assumption is similar to the assumption  used in the known  KLT-like transforms.

\subsection{Problem discussion}\label{diss}
We note that (\ref{f1j}) and (\ref{axj}) can be represented as
\begin{eqnarray}\label{ffjj}
\f_j[\y(t_j,\cdot)]=\x(t_j,\cdot).
\end{eqnarray}
Also, in (\ref{min1}), the term $\dr\x(t_j,t_{j+1},\cdot) - \g_j [\dr\y(t_j,t_{j+1},\cdot)]$ can be rewriten as
\begin{eqnarray*}
&&\dr\x(t_j,t_{j+1},\cdot) - \g_j [\dr\y(t_j,t_{j+1},\cdot)]\\
&& =\x(t_{j+1},\cdot)- (\aab_j + \g_j[\y(t_{j+1},\cdot)])\\
&& =\x(t_{j+1},\cdot)-\f_j[\y(t_{j+1},\cdot)].
\end{eqnarray*}
Therefore, (\ref{min1})--(\ref{r1}) can be represented as
\begin{eqnarray}\label{ffjj1}
\f_j[\y(t_{j+1},\cdot)]\approx \x(t_{j+1},\cdot).
\end{eqnarray}
In other words, the relations (\ref{f1j}), (\ref{axj}) and (\ref{min1})--(\ref{r1}) mean that we wish to determine  $\f_j$ so that
$\f_j$ exactly  interpolates $\x(t,\cdot)$ at $t=t_j$ and approximately interpolates $\x(t,\cdot)$ at $t=t_{j+1}$, as in (\ref{ffjj}) and (\ref{ffjj1}), respectively.

By this reason, the pairs of signals  $\{\x(t_1,\cdot), \y(t_1,\cdot)\},$ $\ldots, \{\x(t_{p-1},\cdot), \y(t_{p-1},\cdot)\}$ are called
{\em the interpolation pairs}.

It is worthwhile to note that for the case of pure filtering (with no compression) the constraint (\ref{r1}) is omitted.

\section{Main results} \label{main}

\subsection{Best rank-constrained matrix approximation }\label{rank}

First we recall a recent result on the best rank constrained matrix approximation \cite{tor-f1,fri2} which will be
used in the next section.

 Let $\C^{m\times n}$ be a set of $m\times n$ complex valued matrices, and denote by
 $\cR(m,n,k)\subseteq \C^{m\times n}$ the variety of all $m\times n$ matrices of rank
 $k$ at most.  Fix $A=[a_{ij}]_{i,j=1}^{m,n}\in \C^{m\times n}$. Then $A^*\in\C^{n\times m}$
 is the conjugate transpose of $A$.  Let the SVD of $A$ be given by
  \begin{equation}\label{ausv}
 A=U_A\Sigma_A V_A^*,
 \end{equation}
 where
 $
 U_A\in \C^{m\times m}$ and $V_A\in \C^{n\times n} \quad \mbox{are unitary matrices,}
 $
 $\Sigma_A:=\mbox{diag}(\sigma_1(A),\ldots,\sigma_{\min(m,n)}(A))$ $\in\C^{m\times n}
 $
is a generalized diagonal matrix,  with the singular values $\sigma_1(A)\ge \sigma_2(A)\ge\ldots\ge 0$
 on the main diagonal.

   Let $U_A=[u_1\;u_2\;\ldots u_m]$ and  $V_A=[v_1\;v_2\;\ldots v_n]$
 be the representations of $U$ and $V$ in terms of their $m$ and $n$ columns, respectively.
 Let
 \begin{equation}\label{defpalf}
 P_{A,L}:=\sum_{i=1}^{\rank A} u_iu_i^* \in \C^{m\times m} \hspace*{-3mm}\qa \hspace*{-2mm}
 P_{A,R}:=\sum_{i=1}^{\rank A} v_iv_i^* \in \C^{n\times n}
 \end{equation}
 be the orthogonal projections on the range of $A$ and $A^*$, correspondingly.
 Define
  \begin{equation}
 \label{63-ak-svd}
 A_k:=\langle \langle  A\rangle\rangle_k:=\sum_{i=1}^k \sigma_i(A)u_i v_i^* = U_{Ak}\Sigma_{Ak} V_{Ak}^*\in \C^{m\times n}
\end{equation}
  for $k=1,\ldots,\rank A$, where
 \begin{eqnarray}\label{63-usv-k}
 &&\hspace*{-13mm}U_{Ak}= [u_1\;u_2\;\ldots u_k], \hspace*{-2mm}\quad \Sigma_{Ak}=\mbox{diag}(\sigma_1(A),\ldots,\sigma_k(A))\\
 && \qa
 V_{Ak}=[v_1\;v_2\;\ldots v_k].
\end{eqnarray}

 For $k>\rank A,$ we write $A_k:=A\;(=A_{\rank A})$. For $1\le k<\rank A$, the matrix $A_k$ is uniquely defined
 if and only if $\sigma_k(A)>\sigma_{k+1}(A)$.

  Recall that
$A^{\dagger}:=V_A\Sigma_A^{\dagger}U_A^*\in \C^{n\times m}$ is the Moore-Penrose generalized inverse of $A$,
 where
 \noindent
 $\displaystyle \Sigma_A^{\dagger}:=\diag \left(\frac{1}{\sigma_1(A)},\ldots,\frac{1}{\sigma_{\rank A}(A)},0,\ldots,0\right)
 \in \C^{n\times m}$.  See for example \cite{ben1}.

Henceforth $\|\cdot\|$ designates the Frobenius norm.

Theorem \ref{imprthm2} below provides a solution to the problem of finding a matrix $X_0$ such that
 \begin{equation}\label{prob}
||A - BX_0C|| = \min_{X\in \cR(p,q,k)} ||A -BXC||
 \end{equation}
and is based on the fundamental result in \cite{fri2} (Theorem 2.1) which  is a generalization of the well known
Eckart-Young theorem \cite{gol1}. The Eckart-Young theorem states that for the case when
 $m=p$, $q=n$ and $B=I_m,$ $C=I_n$, the solution is given by $X_0=A_k$, i.e.
 \begin{equation}\label{svdap}
 ||A - A_k|| = \min_{X\in \cR(m,n,k)} ||A - X||, \quad k=1,\ldots,\min \{m,n\}.
 \end{equation}

 \begin{theorem}\cite{tor-f1}\label{imprthm2} Let $A\in \C^{m\times n},$ $B\in \C^{m\times p}$ and
 $C\in \C^{q\times n}$ be given matrices. Let
\begin{equation}\label{pq}
L_B:=(I_p-P_{B,R})S \qa L_C:=T(I_q-P_{C,L})
\end{equation}
where   $S\in \C^{p\times p}$ and $T\in \C^{q\times q}$ are any matrices, and $I_p$ is the $p\times p$ identity matrix.
Then the matrix
\begin{equation}\label{xsolform}
X_0:=(I_p+L_B)B^{\dagger}\langle\langle P_{B,L}AP_{C,R}\rangle\rangle_k C^{\dagger}(I_q+L_C)
\end{equation}
is a minimizing matrix for the minimal problem $(\ref{prob})$. Any minimizing $X_0$ has the above form.
\end{theorem}

\subsection{Determination of the transform ${F}$ }\label{det1}

Let us introduce the inner product
\begin{eqnarray}\label{w3}
&&\hspace*{-10mm} \left\langle \dr \x^{(i)}(t_j,t_{j+1},\cdot), \dr \y^{(k)}(t_j,t_{j+1},\cdot)\right\rangle \nonumber\\
&&\hspace*{-3mm}= \int_{\Omega} \dr \x^{(i)}(t_j,t_{j+1},\omega) \dr \y^{(k)}(t_j,t_{j+1}, \omega)\ d\mu (\omega)
\end{eqnarray}
and the covariance matrix
$$
E_{\dr x_j \dr y_j} = \left \{\left\langle \dr \x^{(i)}(t_j,t_{j+1},\cdot), \dr \y^{(k)}(t_j,t_{j+1},\cdot)\right\rangle
\right \}_{i,k=1}^{m, n}.
$$
We denote by $X^{1/2}$ a matrix such that $X^{1/2}X^{1/2} = X$.

\begin{theorem}\label{thm2}  The proposed transform ${F}$ is given by
\begin{equation}\label{fyd0}
{F}[\y(t,\omega)] = \sum_{j=1}^{p-1}\delta_j  {F}_j [\y(t,\omega)]
\end{equation}
where
\begin{eqnarray}\label{f0jxj}
{F}_j [\y(t,\omega)] =  \x(t_j,\omega)  + {G}_j[\y(t,\omega) - \y(t_j,\omega)],
\end{eqnarray}
\begin{eqnarray}\label{b0j}
&&\hspace*{-13mm}{G}_j =
\langle\langle E_{\dr x_{j}\dr y_{j}}(E_{\dr y_{j}\dr y_{j}}^{1/2})^{\dag}\rangle\rangle_{r_j} (E_{\dr y_{j}\dr y_{j}}^{1/2})^{\dag}
\nonumber\\
&&\hspace*{13mm}+ M_{Gj}[I_n- E_{\dr y_{j}\dr y_{j}}^{1/2} (E_{\dr y_{j}\dr y_{j}}^{1/2})^{\dag}],
\end{eqnarray}
and  $M_{Cj}\in\rt^{m\times n}$ is an arbitrary matrix.
\end{theorem}

\begin{proof}  First, we note that  ${F}_j$ in (\ref{f0jxj}) has the form as in (\ref{fyab}) where
\begin{eqnarray}\label{abj1}
 \aab_j(\omega) = \x(t_j,\omega) - {G}_j[\y(t_j,\omega)].
\end{eqnarray}

To find $\g_j$ that satisfies (\ref{min1}) and  (\ref{r1}), we write:
\begin{eqnarray}\label{drx1}
 &&\hspace*{-10mm}\left \|\dr\x(t_j,t_{j+1},\cdot) - \g_j (\dr\y(t_j,t_{j+1},\cdot))\right \|^2_\Omega \nonumber\\
 && \hspace*{-10mm}=  \mbox{tr} \{E_{\dr x_{j} \dr x_{j}} - E_{\dr x_{j}\dr y_{j}}G_j^T - G_j E_{\dr y_{j}\dr x_{j}}
+ G_j E_{\dr y_{j}\dr y_{j}} G_j^T\}\nonumber \\
\label{62-er2} & = &\|E_{\dr x_{j}\dr x_{j}}^{1/2}\|^2 -
\|E_{\dr x_{j}\dr y_{j}}(E_{\dr y_{j}\dr y_{j}}^{1/2})^\dag\|^2\nonumber \\
&& \hspace*{20mm} + \|(G_j - E_{\dr x_{j}\dr y_{j}}E_{\dr y_{j}\dr y_{j}}^{\dag}) E_{\dr y_{j}\dr y_{j}}^{1/2}\|^2 \nonumber \\
& = &\|E_{\dr x_{j}\dr x_{j}}^{1/2}\|^2 - \|E_{\dr x_{j}\dr y_{j}}(E_{\dr y_{j}\dr y_{j}}^{1/2})^\dag\|^2 \nonumber \\
&& \hspace*{15mm}+ \|E_{\dr x_{j}\dr y_{j}}(E_{\dr y_{j}\dr y_{j}}^{1/2})^{\dag} - G_jE_{\dr y_{j}\dr y_{j}}^{1/2}\|^2.
\end{eqnarray}
The latter is true because
$$
E_{\dr y_{j}\dr y_{j}}^{\dag} E_{\dr y_{j}\dr y_{j}}^{1/2} = (E_{\dr y_{j}\dr y_{j}}^{1/2})^{\dag}
$$
and
\begin{equation}\label{62-ee}
E_{\dr x_{j}\dr y_{j}}E_{\dr y_{j}\dr y_{j}}^\dag E_{\dr y_{j}\dr y_{j}} = E_{\dr x_{j}\dr y_{j}}
\end{equation}
by Lemma 24 in \cite{tor100}.

In (\ref{62-er2}), the only term that depends on $G_j$ is $\|E_{\dr x_{j}\dr y_{j}}(E_{\dr y_{j}\dr y_{j}}^{1/2})^{\dag}
- G_jE_{\dr y_{j}\dr y_{j}}^{1/2}\|^2$. Thus a determination of $G_j$ is reduced to the problem (\ref{prob}) with
\begin{eqnarray}\label{aex1}
&&A = E_{\dr x_{j}\dr y_{j}}(E_{\dr y_{j}\dr y_{j}}^{1/2})^{\dag}, \hspace*{1mm} B=I_n, \hspace*{1mm} X=G_j \\
&&\hspace*{19mm} \mbox{and} \hspace*{3mm} C = E_{\dr y_{j}\dr y_{j}}^{1/2}.\nonumber
\end{eqnarray}
Let the SVD of $E_{\dr y_{j}\dr y_{j}}^{1/2}$ be given by
\begin{equation}\label{eedr}
E_{\dr y_{j}\dr y_{j}}^{1/2}=V_n\Sigma V_n^T
\end{equation}
 where $V_n=[v_1,\ldots, v_n]$ with $v_i$ the $i$-th column of $V_n,$
$\Sigma = \mbox{diag}(\sigma_1,\ldots,\sigma_{n})\in\rt^{n\times n}$ and $\sigma_1,\ldots,\sigma_{n}$ are associated
singular values. Let $\rank E_{yy}^{1/2}=\rho$. In this case, the solution by Theorem \ref{imprthm2} is given by
\begin{equation}\label{gj1}
G_j = N_j + N_j T_j(I_n - V_\rho V_\rho^T),
\end{equation}
where
\begin{equation}\label{nj1}
N_j =
\langle\langle E_{\dr x_{j}\dr y_{j}}(E_{\dr y_{j}\dr y_{j}}^{1/2})^{\dag}V_\rho V_\rho^T\rangle\rangle_{r_j}
(E_{\dr y_{j}\dr y_{j}}^{1/2})^{\dag}
\end{equation}
 and $T_j$ is an arbitrary matrix. Here,
 \begin{equation}\label{edryj1}
(E_{\dr y_{j}\dr y_{j}}^{1/2})^{\dag}V_\rho V_\rho^T = V_\rho \Sigma_\rho^{-1} V_\rho^T  V_\rho V_\rho^T
=(E_{\dr y_{j}\dr y_{j}}^{1/2})^{\dag}
 \end{equation}
 and $\Sigma_\rho^{-1}= \diag(\frac{1}{\sigma_1},\ldots,\frac{1}{\sigma_{\rho}})$.
 If we choose $M_{Cj} = N_j T_j$ then (\ref{b0j}) follows from (\ref{gj1})--(\ref{edryj1}) because
$V_\rho V_\rho^T = E_{\dr y_{j}\dr y_{j}}^{1/2} (E_{\dr y_{j}\dr y_{j}}^{1/2})^{\dag}$.  
\end{proof}

\subsection{Procedure of compression and de-compression}\label{part}

For each $t\in T$, we wish to filter the observed data $\y(t,\cdot)\in L^{2}(\Omega,{\mathbb R}^{n})$, compress it to a shorter vector
$\widetilde{\x}(t,\cdot)\in L^{2}(\Omega,{\mathbb R}^{r_j})$ with $r_j=\min\{m, n\}$ components\footnote{Recall
that in statistics those components are called the principal components \cite{jol1}.} and then to
reconstruct $\widetilde{\x}(t,\cdot)$ in the form $\widehat{\x}(t,\cdot)$ so that $\widehat{\x}(t,\cdot)$ is close to $\x(t,\cdot)$.

This procedure  is provided by the proposed transform $F$ for two special cases,

(i) $t\in (t_1, \hspace*{1mm} b]$, and

(ii) $t=t_1$

\noindent
as follows. We consider the cases successively.

(i) For $t\in (t_1, \hspace*{1mm} b]$, compression and filtering of $\y(t,\cdot)$   is performed via a representation of $G_j$ in
(\ref{b0j}) as a product of two matrices,
$$
G_j = D_{G_j} C_{G_j},
$$
where $D_{G_j}$ is a $m\times r_j$ matrix and $C_{G_j}$ is a $r_j\times n$ matrix. Then matrix $C_{G_j}$ compresses
 $\x(t_j,\omega)$ to a vector with $r_j$ components and matrix $D_{G_j}$ reconstructs the compressed vector so that the
 reconstruction is close to $\x(t_j,\omega)$.

Matrices $D$ and $C_j$ are determined as follows.

Let us write the SVD of $G_j$ in (\ref{b0j}) as
 \begin{equation}\label{svd-ej}
U_{G_j}\Sigma_{G_j} V_{G_j}^T =G_j
\end{equation}
where matrices
 \begin{eqnarray*}
&& U_{G_j}=[u_{j1},\ldots, u_{jm}]\in\rt^{m\times m},\\
&& \Sigma_{G_j}=\mbox{diag}(\sigma_{1}(G_j),\ldots, \sigma_{\min(m,n)}(G_j))\in\rt^{m\times n}\\
&&\hspace*{-12mm}\qa V_{G_j}=[v_{j1},\ldots, v_{jn}]\in\rt^{n\times n}
 \end{eqnarray*}
are similar to matrices $U_A$, $\Sigma_A$ and $ V_A$ for the SVD of matrix $A$ in (\ref{ausv}), respectively. In particular,
 $\sigma_{j1}$, $\ldots,$ $\sigma_{j \min(m,n)}$ are the associated singular values.

Let us denote
 \begin{eqnarray}\label{ugjrj1}
&& U_{G_j r_j}=[u_{j1},\ldots, u_{j r_j}] \in\rt^{m\times r_j},\\\label{ugjrj2}
&&\Sigma_{G_j r_j}=\mbox{diag}(\sigma_{1}(G_j),\ldots, \sigma_{r_j}(G_j))\in\rt^{r_j\times r_j}\\\label{ugjrj3}
&&\hspace*{-12mm}\qa V_{G_j r_j}=[v_{j1},\ldots, v_{j r_j}]\in\rt^{n\times r_j},
 \end{eqnarray}
  where $r_j$ is as in (\ref{r1}).

Then the transform $\f$ in (\ref{fyd0})--(\ref{b0j}) can be written as
\begin{equation}\label{fyt1}
\f[\y(t,\cdot)] = \sum_{j=1}^{p-1}\delta_j  \f_j [\y(t,\cdot)],
\end{equation}
where
 \begin{equation}\label{gjbb}
F_j[\y(t,\omega)]  = \zz_j(t_j,\omega) +D_{G_j} C_{G_j}\y(t,\omega),
\end{equation}
 \begin{eqnarray}\label{zzj}
 &&\zz_j(t_j,\omega) =\x(t_j,\omega) - G_j\y(t_j,\omega),\\\label{ccj1}
&&\hspace*{-12mm}  D_{G_j} =U_{G_j r_j}\Sigma_{G_j  r_j}\in \rt^{m\times r_j},\hspace*{1mm}C_{G_j}
=V_{G_j r_j}^T\in \rt^{r_j\times n},
 \end{eqnarray}
 or
 \begin{eqnarray}\label{ccj2}
 D_{G_j} =U_{G_j r_j}\in \rt^{m\times r_j}, \hspace*{1mm}
C_{G_j} =\Sigma_{G_j  r_j}V_{G_j r_j}^T\in \rt^{r_j\times n}.
 \end{eqnarray}
In particular, for $t=t_{j+1}$ and $j=1,\ldots,p-1$,
 $$
 {F}_j [\y(t_{j+1},\omega)] =  \zz_j(t_j,\omega) +D_{G_j} C_{G_j}\y(t_{j+1},\omega).
 $$

Thus, for all $t\in (t_1, \hspace*{1mm} b]$, $C_{G_j}$ {\em compresses  $\y(t,\omega)$ to a shorter vector} $\widetilde{\x}(t,\omega)=C_{G_j}\y(t,\omega)$ with $r_j$
components.

{\em The reconstruction} $\widehat{\x}(t,\cdot) = \f[\y(t,\omega)]$ of the reference signal
$\x(t,\omega)$ from the compressed data $\widetilde{\x}(t,\omega)$ is performed by $\f_j$ in (\ref{gjbb}) so that
 \begin{equation}\label{fjb1}
\widehat{\x}(t,\omega)  = \zz_j(t_j,\omega) +D_{G_j} \widetilde{\x}(t,\omega).
\end{equation}

(ii) For $t=t_1$, the sub-transform $F_1$ in (\ref{f0jxj}) is reduced to
$$
{F}_1 [\y(t_1,\omega)] =  \x(t_1,\omega)
$$
that does not provide compression.
To provide compression of $\x(t_1,\omega)$, we represent $F_1$, for $t\in [t_1,\hspace*{1mm} t_2]$, as
\begin{equation}\label{fy1}
 {F}_1 [\y(t,\omega)] = \x(t_2,\omega)  - G_1[\y(t,\omega)-\y(t_2,\omega)].
\end{equation}
The latter follows when, for $t\in [t_1,\hspace*{1mm} t_2]$, condition (\ref{f1j})  is replaced by
\begin{equation}\label{a11}
\aab_1 + \g_1[\y(t_1,\cdot)] = \x(t_2,\cdot),
\end{equation}
and condition (\ref{min1})--(\ref{r1}), for $t\in [t_1,\hspace*{1mm} t_2]$, remains the same. Then, for $t=t_1$, (\ref{fy1})
implies
 \begin{equation}\label{f1yt}
F_1[\y(t_1,\omega)]  = \widetilde{\zz}_1(t_2,\omega) +D_{G_1} C_{G_1}\y(t_1,\omega),
\end{equation}
where
$$
\widetilde{\zz}_1(t_2,\omega) =\x(t_2,\omega)-G_1\y(t_2,\omega),
$$
and $D_{G_1}$ and  $C_{G_1}$ are defined by (\ref{ccj1})--(\ref{ccj2}). Here, $C_{G_1}$ and $D_{G_1}$  perform compression of
$\y(t_1,\omega)$ and subsequent reconstruction as above.

Thus,  by transform $\f$, the  compression of $\y(t,\omega)$ requires $p-1$ matrices $C_j$ and the reconstruction requires
$p-1$ vectors $\zz_j(t_j,\omega)$ and matrices $D_j$.

The compression ratio of transform $\f$ in (\ref{fyt1})--(\ref{gjbb})  is given by
\begin{equation}\label{crj}
c = \frac{r_j}{m} \quad \mbox{where $j=1,\ldots, p-1$}.
\end{equation}
If $r=r_j$ for all $j=1,\ldots, p-1$ then
\begin{equation}\label{cr}
c = \frac{r}{m}.
\end{equation}

\subsection{Error associated with the transform $\f$ in (\ref{fyd0})--(\ref{b0j})}\label{error}
 Here, the  analysis of the error associated with the transform $\f$ in (\ref{fyd0})--(\ref{b0j}) is provided.

Let us introduce the norm by
$$
\|\x(t,\cdot)\|^2_{T,\Omega} = \frac{1}{b-a}{\int}_{T} \|\x(t,\cdot)\|^2_{\Omega} dt.
$$

For $A = E_{\dr x_{j}\dr y_{j}}(E_{\dr y_{j}\dr y_{j}}^{1/2})^{\dag}$, let $\rank A=\ell$, and let
\begin{eqnarray}\label{sigm}
\sigma_{1}\geq \sigma_{2}\geq \ldots \geq\sigma_{\ell}
\end{eqnarray}
 be singular values of  $A$.

The following theorem establishes relationships for the errors associated with the proposed transform $\f$.

\begin{theorem}\label{thm3}
Let transform $\f$ be given by (\ref{fyt1})--(\ref{f1yt}). Let $\widetilde{\x}(t,\cdot)=C_{G_j}\y(t,\cdot)\in L^2(\Omega,\rt^{r_j})$
be the vector  compressed and filtered from $\y(t,\cdot)\in L^2(\Omega,\rt^{n})$ by $C_{G_j}$ in (\ref{ccj1}), (\ref{ccj2}),
where $r_j\leq \min\{m, n\}$. Then the reference signal $\x(t,\cdot)\in L^2(\Omega,\rt^{m})$ is reconstructed
from $\widetilde{\x}(t,\cdot)$ by $\f$ as $\widehat{\x}(t,\cdot) = \f[\y(t,\omega)]$ with any given accuracy in the following
sense:
\begin{eqnarray}\label{ertf}
&&\|\x(t,\cdot) - \f[\y(t,\cdot)]\|^2_{T,\Omega} \rightarrow 0 
\end{eqnarray}
as
\begin{eqnarray}\label{ertj}
&&\hspace*{-13mm}\max_{j=1,\ldots p-1}  \|\x(t,\cdot) - \x(t_j,\cdot)\|^2_{T,\Omega}\rightarrow 0,\\
&&\hspace*{-13.5mm}\max_{j=1,\ldots p-1}  \|\y(t,\cdot) - \y(t_j,\cdot)\|^2_{T,\Omega}\rightarrow 0\\\label{pinf}
&&\hspace*{2mm}\mbox{and}\hspace*{3mm} p\rightarrow \infty.
\end{eqnarray}

For $t=t_{j+1}$ and $j=0,\ldots,p-1$, the error associated with $\f [\y(t_{j+1},\cdot)]$ is given by
\begin{eqnarray}\label{ertfj}
\|\x(t_{j+1},\cdot) - \f [\y(t_{j+1},\cdot)] \|^2_\Omega =\|E_{\dr x_{j}\dr x_{j}}^{1/2}\|^2 -\sum_{i=1}^{r_j} \sigma_i^2.
\end{eqnarray}
where $\sigma_1,\ldots, \sigma_{r_j}$ are defined by (\ref{sigm}).
\end{theorem}

\begin{remark}
Although, by the assumption, $\x(t_{j+1},\cdot)$ is given, its compression implies the associated error presented by (\ref{ertfj}).
\end{remark}

\begin{proof} Let us first show that the statement in (\ref{ertf})--(\ref{ertj}) is true. We have
\begin{eqnarray}\label{}
&&\hspace*{-8mm} \|\x(t,\cdot) - \f [\y(t,\cdot)]\|^2_{T,\Omega}\nonumber\\
&&\hspace*{-8mm}= \|\sum_{j=1}^{p-1}\delta_j [\x(t,\cdot)- \f_j [\y(t,\cdot)]\|^2_{T,\Omega}\nonumber\\
&&\hspace*{-8mm} =\|\sum_{j=1}^{p-1}\delta_j [\x(t,\cdot) - \x(t_j,\cdot)]\nonumber\\
&&\hspace*{17mm} + \hspace*{1mm}G_j [\y(t_j,\cdot)-\y(t,\cdot)]\|^2_{T,\Omega}\nonumber\\
&&\hspace*{-8mm} \leq\max_{j=1,\ldots p-1}\left\{ \|\x(t,\cdot) - \x(t_j,\cdot)\|^2_{T,\Omega}\right.\nonumber\\ \label{max1}
&&\hspace*{17mm}\left.+ \|G_j\|\|\y(t,\cdot)- \y(t_j,\cdot)\|^2_{T,\Omega}\right\}
\end{eqnarray}
where  $G_j$ is given by ((\ref{b0j}).  Then (\ref{ertj})--(\ref{pinf}) and (\ref{max1})
imply (\ref{ertf}).\footnote{In particular, $\oo^\dag = \oo$.}

Let us now consider (\ref{ertfj}). In the notation (\ref{63-ak-svd}) and (\ref{aex1}),
$$
A_{r_j}:= \langle\langle E_{\dr x_{j}\dr y_{j}}(E_{\dr y_{j}\dr y_{j}}^{1/2})^{\dag}\rangle\rangle_{r_j}.
$$
By Lemma 42 in \cite{tor100},
$$
A_{r_j}(E_{\dr y_{j}\dr y_{j}})^{\dag}E_{\dr y_{j}\dr y_{j}} = A_{r_j}.
$$
It is also true \cite{tor100} that
$$
(E_{\dr y_{j}\dr y_{j}}^{1/2})^{\dag} = E_{\dr y_{j}\dr y_{j}}^{\dag}E_{\dr y_{j}\dr y_{j}}^{1/2}.
$$
Therefore, in (\ref{b0j}), $G_j$ can be represented as
\begin{eqnarray*}
&&\hspace*{-15mm}G_j=A_{r_j}(E_{\dr y_{j}\dr y_{j}}^{1/2})^{\dag}
+ M_j[I_n- E_{\dr y_{j}\dr y_{j}}^{1/2} (E_{\dr y_{j}\dr y_{j}}^{1/2})^{\dag}]\\
&&\hspace*{-10mm}=A_{r_j}E_{\dr y_{j}\dr y_{j}}^{\dag}E_{\dr y_{j}\dr y_{j}}^{1/2}\\
&&\hspace*{10mm}+ M_j[I_n- E_{\dr y_{j}\dr y_{j}}^{1/2} (E_{\dr y_{j}\dr y_{j}}^{1/2})^{\dag}].
\end{eqnarray*}
On the basis of (\ref{min1}), (\ref{ffjj1}), (\ref{b0j}) and (\ref{62-er2}), we have
\begin{eqnarray}\label{}
&&\hspace*{-15mm}\|\x(t_{j+1},\cdot) - F [\y(t_{j+1},\cdot)] \|^2_\Omega \nonumber\\
&& = \left \|\dr\x(t_j,t_{j+1},\cdot) - \g_j (\dr\y(t_j,t_{j+1},\cdot))\right \|^2_\Omega \nonumber\\
&& =\|E_{\dr x_{j}\dr x_{j}}^{1/2}\|^2 - \|E_{\dr x_{j}\dr y_{j}}(E_{\dr y_{j}\dr y_{j}}^{1/2})^\dag\|^2 \nonumber \\
\label{er1-1}
&& \hspace*{15mm}+ \|E_{\dr x_{j}\dr y_{j}}(E_{\dr y_{j}\dr y_{j}}^{1/2})^{\dag} - A_{r_j}\|^2
\end{eqnarray}
where \cite{gol1}
\begin{eqnarray}\label{er1-2}
\|E_{\dr x_{j}\dr y_{j}}(E_{\dr y_{j}\dr y_{j}}^{1/2})^{\dag} - A_{r_j}\|^2 = \sum_{i=r_j+1}^\ell \sigma_i^2.
\end{eqnarray}
Since
\begin{eqnarray}\label{er1-3}
\|E_{\dr x_{j}\dr y_{j}}(E_{\dr y_{j}\dr y_{j}}^{1/2})^\dag\|^2=\sum_{i=1}^\ell \sigma_i^2,
\end{eqnarray}
then (\ref{er1-1})--(\ref{er1-3}) imply (\ref{ertfj}).
\end{proof}

\subsection{A particular case: signals are given by their samples. Numerical realization of transform $\f$}\label{part11}

In practice, signals $\x(t,\cdot)$ and $\y(t,\cdot)$ are given by their samples
\begin{eqnarray}\label{xt}
&&X(t)= [\x (t,\omega_1) \ldots  \x (t,\omega_q)]\in \rt^{m\times q} \\ \label{ytj}
&&\hspace*{-6.5mm}\mbox{and}\hspace*{2mm}
Y(t)= [\y (t,\omega_1) \ldots  \y (t,\omega_q)]\in \rt^{n\times q}.
\end{eqnarray}
respectively. In particular, for $t=t_j$, the samples are
\begin{eqnarray}\label{xtj}
&&\hspace*{-3mm}X_j:=X(t_j)= [\x (t_j,\omega_1) \ldots  \x (t_j,\omega_q)]\in \rt^{m\times q} \\ \label{ytj}
&&\hspace*{-9.5mm}\mbox{and}\hspace*{2mm}
Y_j:=Y(t_j)= [\y (t_j,\omega_1) \ldots  \y (t_j,\omega_q)]\in \rt^{n\times q}.
\end{eqnarray}
In this case, the transform $\f$ in (\ref{fyd})--(\ref{fydj}) takes the form
\begin{equation}\label{fyd1}
F[Y(t,\cdot)] = \sum_{j=1}^{p-1}\delta_j  F_j [Y(t)],
\end{equation}
 where
 \begin{equation}\label{fydj1}
 F_j [Y(t)] = A_j + G_jY(t)\hspace*{2mm}\mbox{and}\hspace*{2mm}  \delta_j = \left\{ \begin{array}{cl}
                             1,  & \hspace*{-2mm}\mbox{if $t_j\leq t \leq t_{j+1}$},\\
                             0, & \hspace*{-2mm}\mbox{otherwise,}
                              \end{array} \right.
\end{equation}
and $A_j=[\aab_j(\omega_1),\ldots,\aab_j(\omega_q)]\in\rt^{m\times q}$ and $G_j\in\rt^{m\times n}$ are matrices such that,
for $j=1,\ldots,p-1$, $A_j$ satisfies the condition
\begin{equation}\label{aa1}
A_j + G_jY_j= X_j
\end{equation}
and $G_j$ solves
\begin{equation}\label{gg1}
\min_{G_j} \|\dr X_j - G_j \dr Y_j\|
\end{equation}
subject to
\begin{equation}\label{rr1}
\rank G_j = r_j\leq min\{m, n\}.
\end{equation}
The solution is provided by the following Corollary which is a particular case of Theorem \ref{thm2}.

\begin{corollary}\label{cor1} In a practical case when signals $\x(t,\cdot)$ and $\y(t,\cdot)$ are represented by samples
(\ref{xt})--(\ref{ytj}),  the required transform is given by (\ref{fyd1})--(\ref{fydj1}) where
\begin{equation}\label{ffg1}
F_j [Y(t)] = X_j - G_j [Y(t) - Y_j]
\end{equation}
with
\begin{eqnarray}\label{ggj1}
&&\hspace*{-13mm}G_j = \langle\langle \dr X_j\dr Y_j^T [(\dr Y_j\dr Y_j^T)^{1/2}]^{ \dag}\rangle\rangle_{r_j}
[(\dr Y_j\dr Y_j^T)^{1/2}]^{\dag}\nonumber\\
&& \hspace*{-8mm} + M_j[I_n - ((\dr Y_j\dr Y_j^T)^{1/2})^{\dag}(\dr Y_j\dr Y_j^T)^{1/2}].
\end{eqnarray}
where matrix $M_j$ is arbitrary.
\end{corollary}

\subsection{ Numerical realization of transform $\f$}\label{num}

In fact, formulas (\ref{ffg1})--(\ref{ggj1}) represent a numerical realization of the transform presented by
(\ref{fyd0})--(\ref{b0j}). In practice, $M_j$ can be chosen as the zero matrix as it is normally done in the KLT-like techniques.

\subsection{Summary of the proposed transform}\label{summ}

Here, we provide a summary of the transform presented in Sections \ref{det1} and \ref{part}.

Let $K_{_X}=\{ \x(t, \cdot) \in L^{2}(\Omega,{\mathbb R}^{m})\hspace*{1mm}| \hspace*{1mm} t \in T\}$ and
$K_{_Y}=\{ \y(t, \cdot) \in L^{2}(\Omega,{\mathbb R}^{n})\hspace*{1mm}| \hspace*{1mm} t \in T\}$ be infinite sets of reference
signals and observable signals, respectively (see  Section \ref{some}).

Choose finite subsets of $K_{_X}$ and $K_{_Y}$,  $\{\x(t_{1},\cdot),$ $\ldots,$ $\x(t_{p-1},\cdot)\}$ and $\{\y(t_{1},\cdot),\ldots, \y(t_{p-1},
\cdot)\}$, respectively.


To process an observable signal $\y(t, \cdot)$ we use the transform $\f$ as presented in (\ref{fyt1}), (\ref{gjbb}) where $D_j$
and $C_j$ are given by  (\ref{ccj1}) or (\ref{ccj2}). The matrices $D_j$ and $C_j$ are constructed from
the SVD (\ref{svd-ej}) truncated to the form (\ref{ugjrj1})--(\ref{ugjrj3}).

The matrix $C_j$ compresses and filters $\y(t, \cdot)$ to a shorter vector $\widetilde{x}(t, \cdot)$ (the vector of principal
components). The reconstruction, $\widehat{x}(t, \cdot)$, is performed in the form (\ref{fjb1}) so that
$\widehat{x}(t, \cdot)$ is close to the reference signal  $\x(t, \cdot)$ in the sense (\ref{ertf})--(\ref{pinf}).

\subsection{Advantages of proposed transform. Comparison with  KLT  and its extensions}\label{comp}


The proposed methodology  provides {\em a single} transform to process any signal from an {\em infinite}  set of signals.
This is {\em a distinctive feature} of the considered technique.

To the best of our knowledge, there is  only one other known approach \cite{tor1} that provides the transform with a
similar property. Moreover, the KLT is a particular case of the transform \cite{tor1}.

 Therefore, it is natural to compare the proposed transform with that in \cite{tor1}.
The transform developed in \cite{tor1}  is presented in the special form of a sum with $p$ terms  where
each term is determined from the preceding terms as a solution of a minimization problem for the associated error.
The final term in such an iterative procedure provides signal compression and decompression. In particular, the KLT \cite{tor100}
follows from \cite{tor1} if $p=1$.

Here, we first compare the  transform in  \cite{tor1} and the proposed transform $\f$,  and
demonstrate the advantages of the transform $\f$. Then we show that the similar advantages  also occur
in comparison with  other transforms  based on the KLT idea.

\subsubsection{Associated errors of transforms ${\f}$ and  {\em \cite{tor1}}} Together with the distinctive feature of the
transform $\f$ mentioned above, its other {\em distinctive  property} is an arbitrarily small associated error in the
reconstruction of the reference signal $\x(t,\cdot)$ (Theorem \ref{thm3} in Section \ref{error}). This is achieved under conditions
(\ref{ertj})--(\ref{pinf}).

The transform in \cite{tor1} does not provide such a nice property.

In other words,  $\f$ given by (\ref{fyt1})--(\ref{fjb1}) is composed
from sequences of signals measured at
different time instants $t_i$, not from the averages of signals over the domain  $T \times \Omega$ as in \cite{tor1}.
Therefore, the transform ${\f}$ is `flexible' to variations of signals $\x(t,\cdot)$ and $\y(t,\cdot)$, and  this  inherent
 feature leads to the decrease in the  associated error as it is established in Theorem \ref{thm3} above.

\subsubsection{Degrees of freedom to reduce the associated  error}\label{degree}

Both the transform  $\f$ given by (\ref{fyt1})--(\ref{fjb1}) and the transform in \cite{tor1} have two degrees of freedom
to reduce associated  error: the number of terms that compose the transform, and the matrix ranks.  At the same time,  the error
associated with
the method in \cite{tor1} is bounded despite the increase in its number of terms  while for the
the proposed transform $\f$, the increase in number of terms leads to an arbitrarily small associated error
(see (\ref{ertf})--(\ref{pinf})). This occurs because of a {\em `flexibility' of the transform $\f$} discussed in the following
Section \ref{numer}.

Moreover, unlike the method in \cite{tor1} the proposed approach may provide one more degree or freedom, a distribution of
interpolation
pairs $\{\x(t_j,\cdot), \y(t_j,\cdot)\}_{j=1}^p$ related to the distribution of
points $t_1,\ldots, t_p$. An `appropriate' selection of $\{\x(t_j,\cdot), \y(t_j,\cdot)\}_{j=1}^p$ may diminish the error.
At the same time, an optimal choice of $\{\x(t_j,\cdot), \y(t_j,\cdot)\}_{j=1}^p$ is a specific problem which is not under consideration in this
paper.

\subsubsection{Associated assumptions. Numerical realization}\label{numer}

A numerical realization of the KLT based transforms requires a knowledge or estimation of related covariance matrices.
The estimates are normally found from samples  $X(t_1)$, $\ldots$, $X(t_p)$ and $Y(t_1)$, $\ldots$,
$Y(t_p)$ presented in Section \ref{part11}.
That is, the {\em assumptions} used in numerical  realizations of the KLT based techniques and
the proposed transform $\f$ (given by (\ref{fyt1})--(\ref{f1yt})) {\em are,
in fact, the same}: it is assumed that the samples  $X(t_1)$, $\ldots$, $X(t_p)$, $Y(t_1)$, $\ldots$, $Y(t_p)$ are known.
In other words, the preparatory work that should be done for the  numerical realization of the KLT and our transform ${\f}$ is the
same. At the same time, { the usage of those signal samples  is different.} While the transform in \cite{tor1}
requires estimates of two covariance matrices in the form of two related {\em  averages} formed from  $X(t_1)$,
$\ldots$, $X(t_p)$, $Y(t_1)$, $\ldots$, $Y(t_p)$, our transform $\f$ uses {\em each representative} $X(t_j)$ and
$Y(t_j)$  separately. This makes  {\em transform $\f$ `flexible' to a variation} of $\x(t,\cdot)$ and $\y(t,\cdot)$.

In the case of finite signal sets, $\{\y_{(1)},\ldots, \y_{(N)}\}$ and
$\{\x_{(1)},\ldots, \x_{(N)}\}$, we might apply the individual KLT transform $W_i$ to each pair $\{\y_{(i)},\x_{(i)}\}$,
for $i=1,\ldots,N$, separately.
Such a procedure would require $N$ times more computational work compared with the transform in \cite{tor1}.

As a result, while computational efforts associated with the proposed technique are similar or less than  those needed
for the KLT based transforms, transform $\f$ given by (\ref{fyt1})--(\ref{ccj2}) allows us to
substantially improve the accuracy in signal estimation, as  has been shown in Theorem \ref{thm3}.
This observation is also demonstrated in the simulations presented in Section \ref{sim} below.

\begin{center}
\begin{tabular}{||c||c|c|c|c||}
\multicolumn{5}{c}{Table 1. Accuracy} \\
\multicolumn{5}{c}{associated with the proposed transform $\f$.}\\
\hline\hline
& \multicolumn{4}{c||} {Number of interpolation pairs}\\
\hline
 & \multicolumn{1}{c|} {$p=9$} &\multicolumn{1}{c|} {$p=21$} &\multicolumn{1}{c|} {$p=41$}& \multicolumn{1}{c||} {$p=81$}\\
\hline\hline
 Compression & \multicolumn{4}{c||} {The best accuracy associated with  }\\
 ratio   & \multicolumn{4}{c||} {our transform $F$, $\varepsilon_{\min} (F)\times 10^{-8}$} \\
\hline\hline
$c=5/116$ &0.7 &0.5 & 0.48 & 0.31\\
\hline
$c=10/116$ & 0.5 &  0.5 & 0.45 &  0.23 \\
\hline\hline
Compression& \multicolumn{4}{c||} {The worst accuracy associated with}\\
 ratio & \multicolumn{4}{c||} {our transform $F$, $\varepsilon_{\max} (F)\times 10^{-8}$} \\
\hline\hline
$c=5/116$ & 1.71& 1.04& 0.71 & 0.64 \\
\hline\hline
$c=10/116$ & 1.49 & 1.07  & 0.76  &  0.56 \\
\hline\hline
\end{tabular}
\end{center}

\vspace*{10mm}

\begin{center}
\begin{tabular}{||c||c|c||c|c||c|c||c|c||}
\multicolumn{9}{c}{Table 2. Comparison in accuracy} \\
\multicolumn{9}{c}{of the proposed transform $\f$ and the KLT.}\\
\hline\hline
& \multicolumn{8}{c||} {Number of interpolation pairs}\\
\hline\hline
 & \multicolumn{2}{c||} {$p=9$} &\multicolumn{2}{c||} {$p=21$} &\multicolumn{2}{c||} {$p=41$}& \multicolumn{2}{c||} {$p=81$}\\
\hline\hline
 Compr. ratio& $\Delta_{{\min}}$ & $\Delta_{{\max}}$ & $\Delta_{{\min}}$ & $\Delta_{{\max}}$  & $\Delta_{{\min}}$ &
 $\Delta_{{\max}}$  & $\Delta_{{\min}}$ & $\Delta_{{\max}}$ \\
\hline\hline
$c=5/116$ & 4.9 & 15. 4 & 9.3 & 18.2&  10.1 & 19.7& 12.4 & 22.9 \\
\hline\hline
$c=10/116$& 4.3 & 17. 4 & 10.3 & 18.7 & 12.9 & 22.0 & 15.7 & 25.1 \\
\hline\hline
\end{tabular}
\end{center}

\begin{center}
\begin{figure}[h]
\centering
 \hspace*{-7mm}\begin{tabular}{c@{\hspace*{5mm}}c}
   \includegraphics[width=8cm,height=2.5cm]{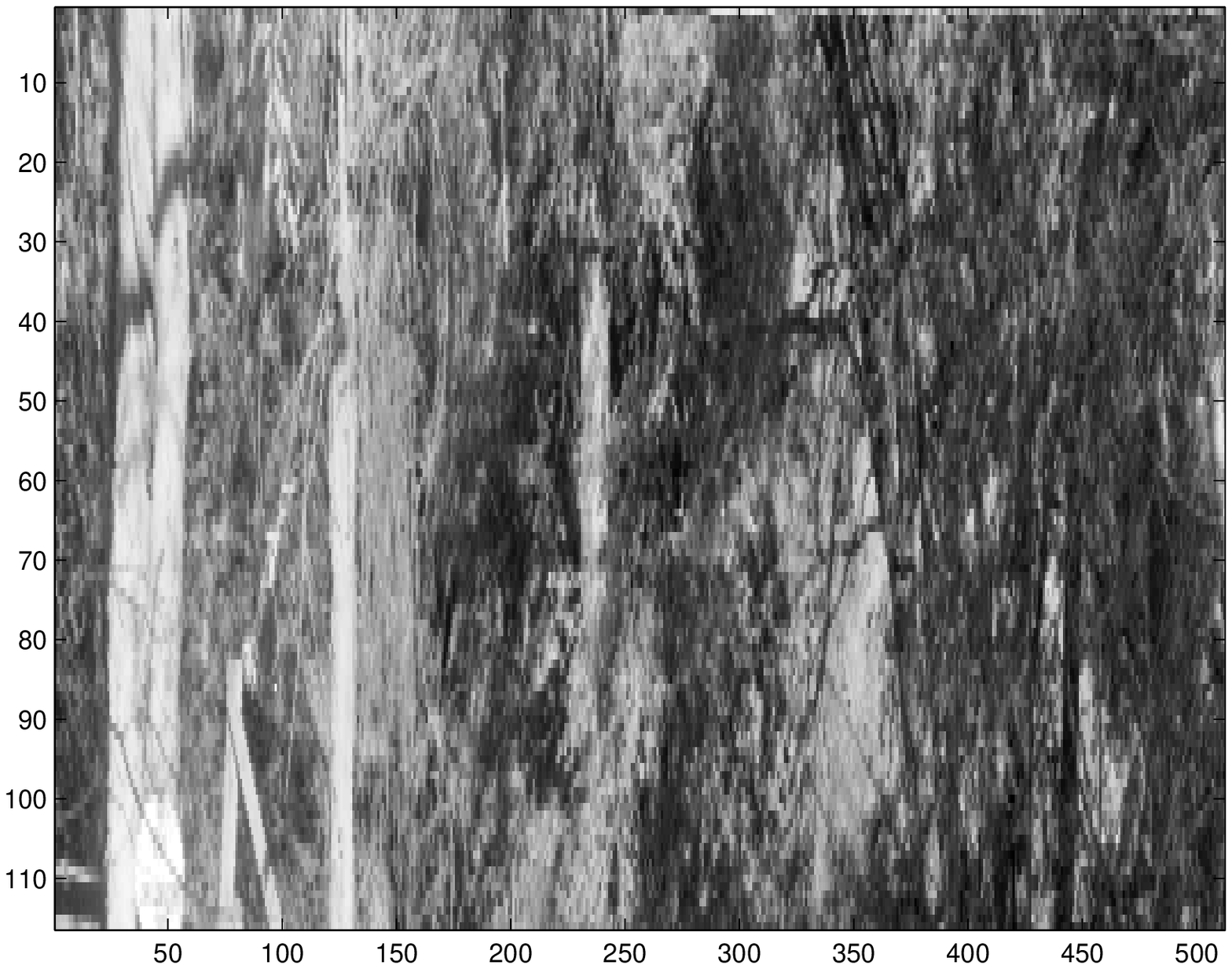} &   \includegraphics[width=8cm,height=2.5cm]{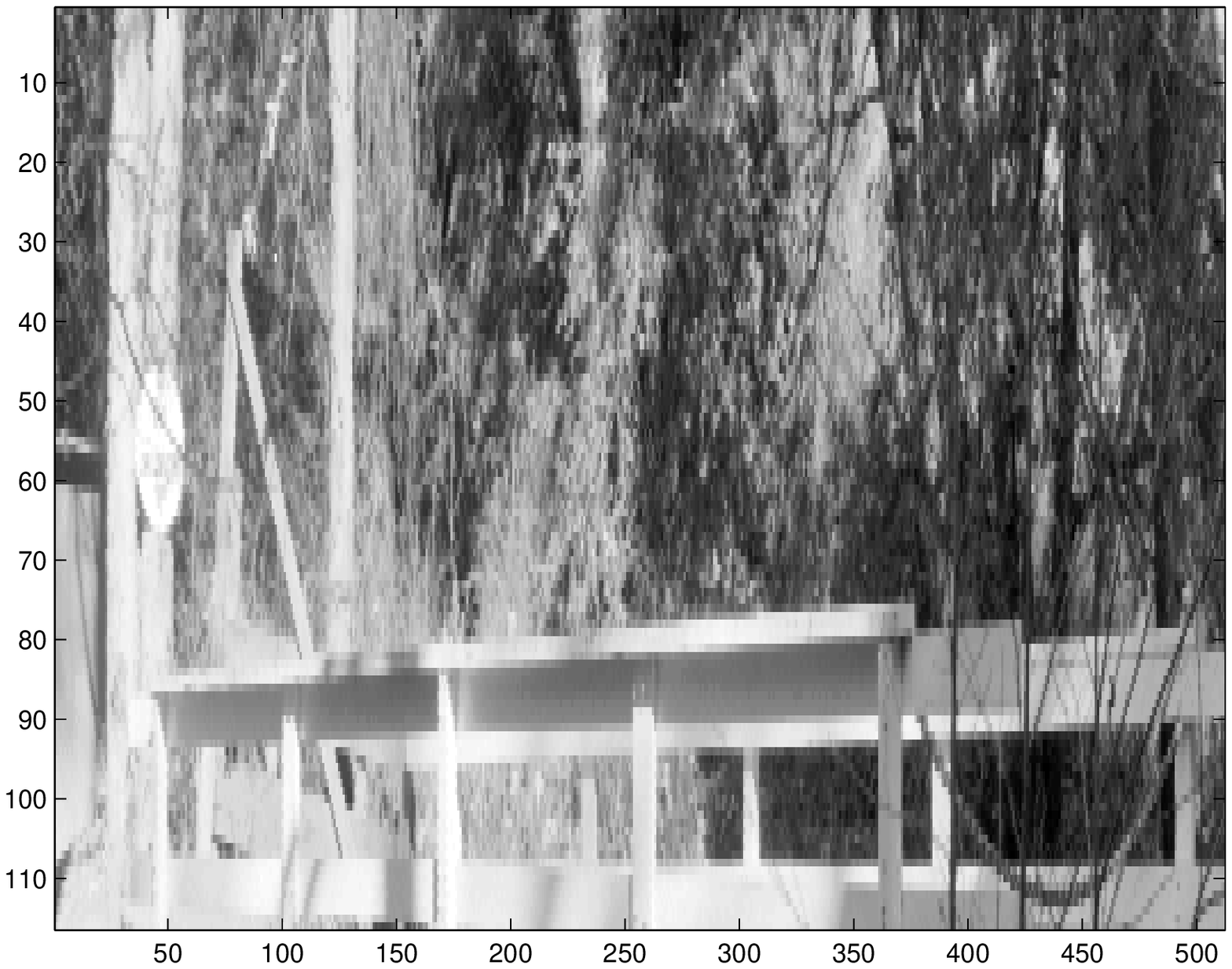} \\
  (a){\small Signal $X^{(1)}.$} &  (b) {\small Signal $X^{(55)}.$}\\
   \includegraphics[width=8cm,height=2.5cm]{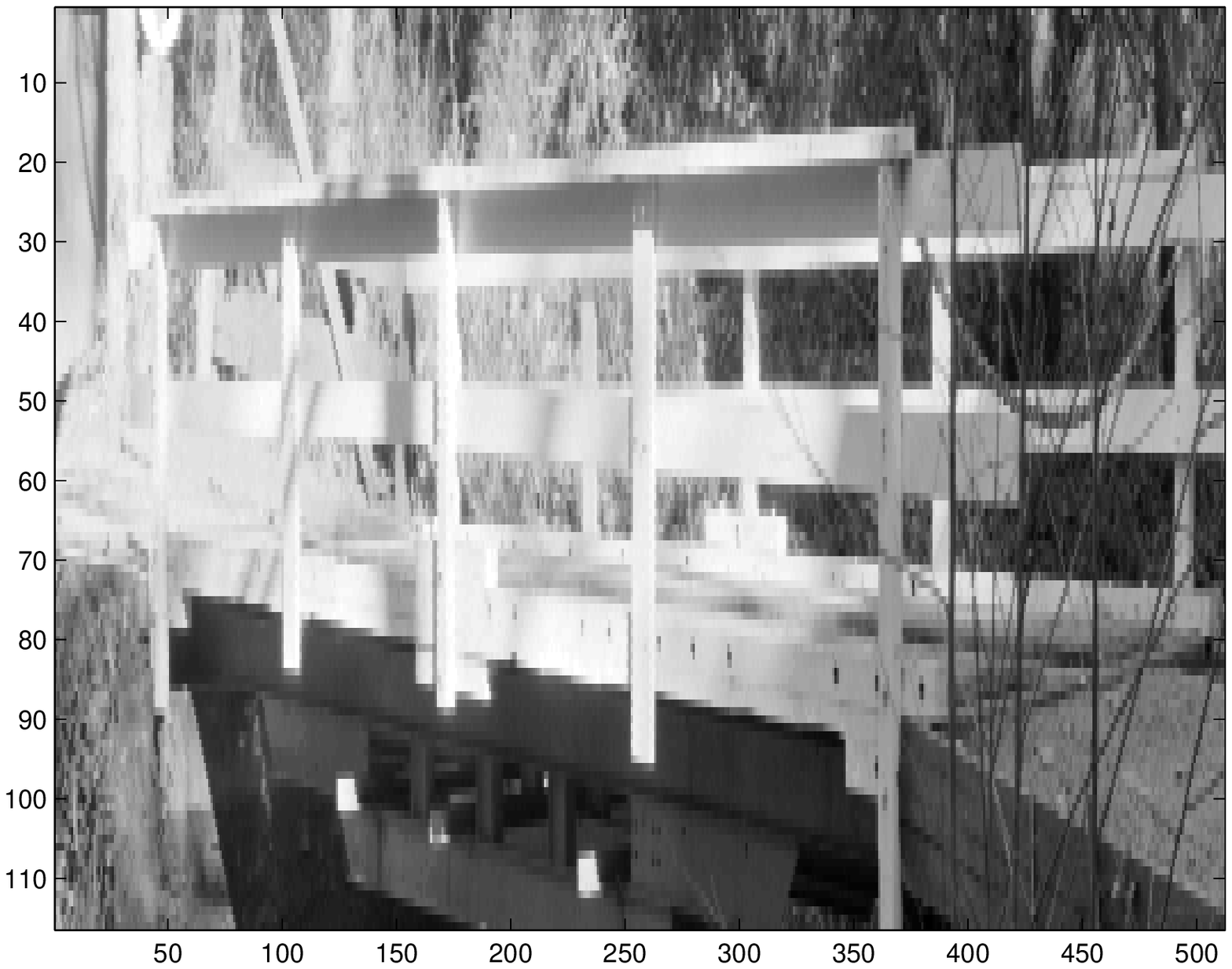} &   \includegraphics[width=8cm,height=2.5cm]{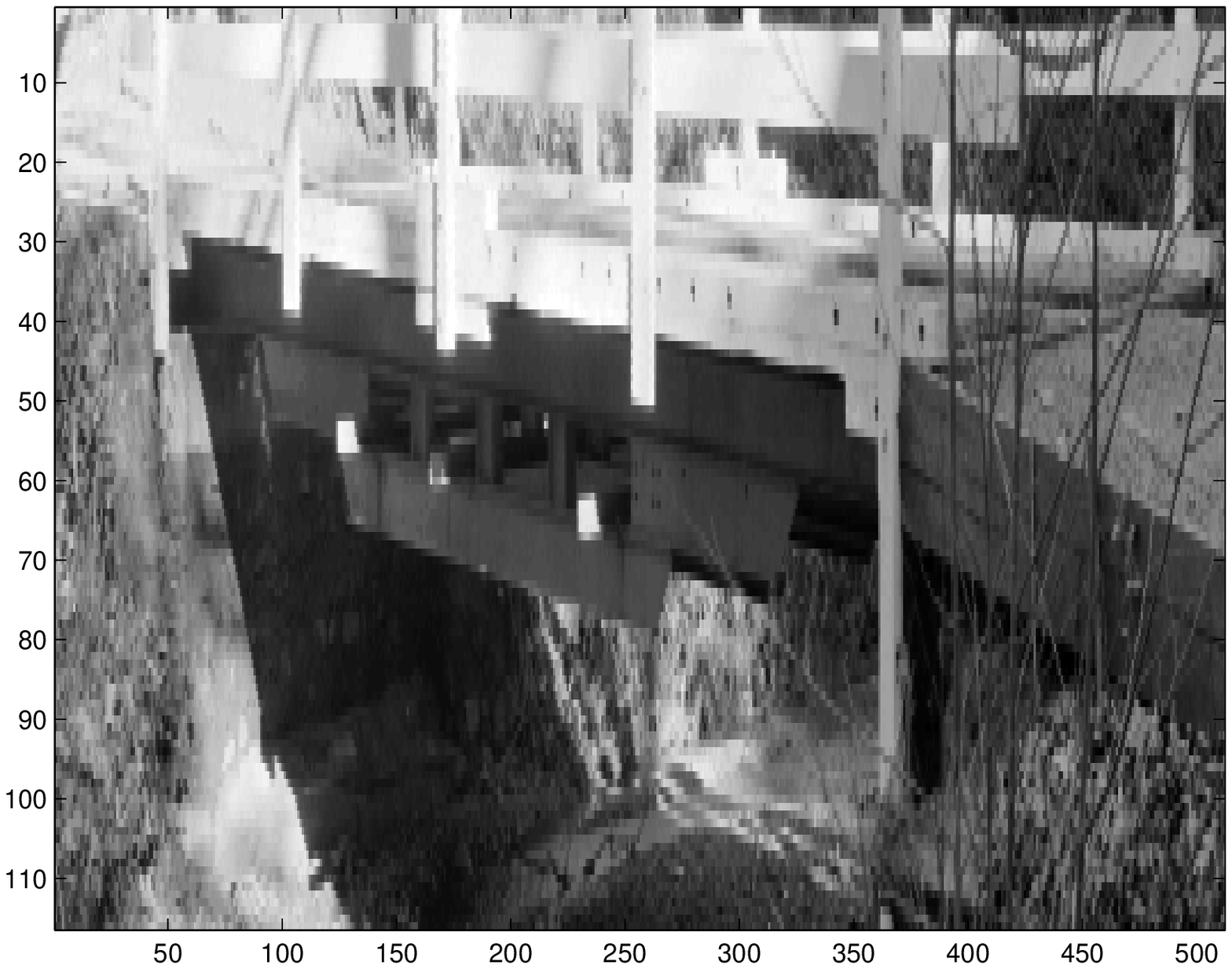} \\
  (c) {\small Signal $X^{(115)}.$} &  (d) {\small Signal $X^{(160)}.$}\\
    \includegraphics[width=8cm,height=2.5cm]{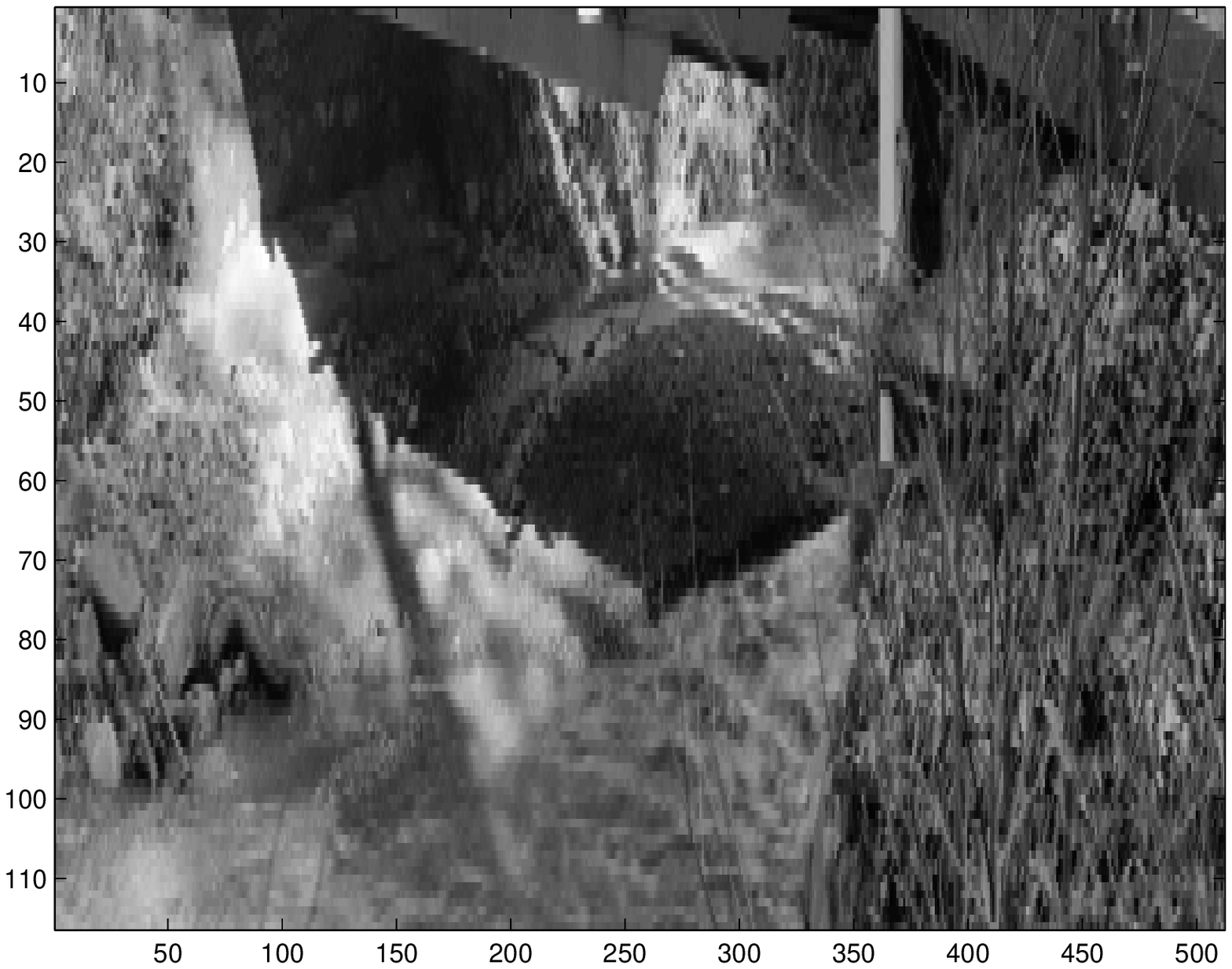} &   \includegraphics[width=8cm,height=2.5cm]{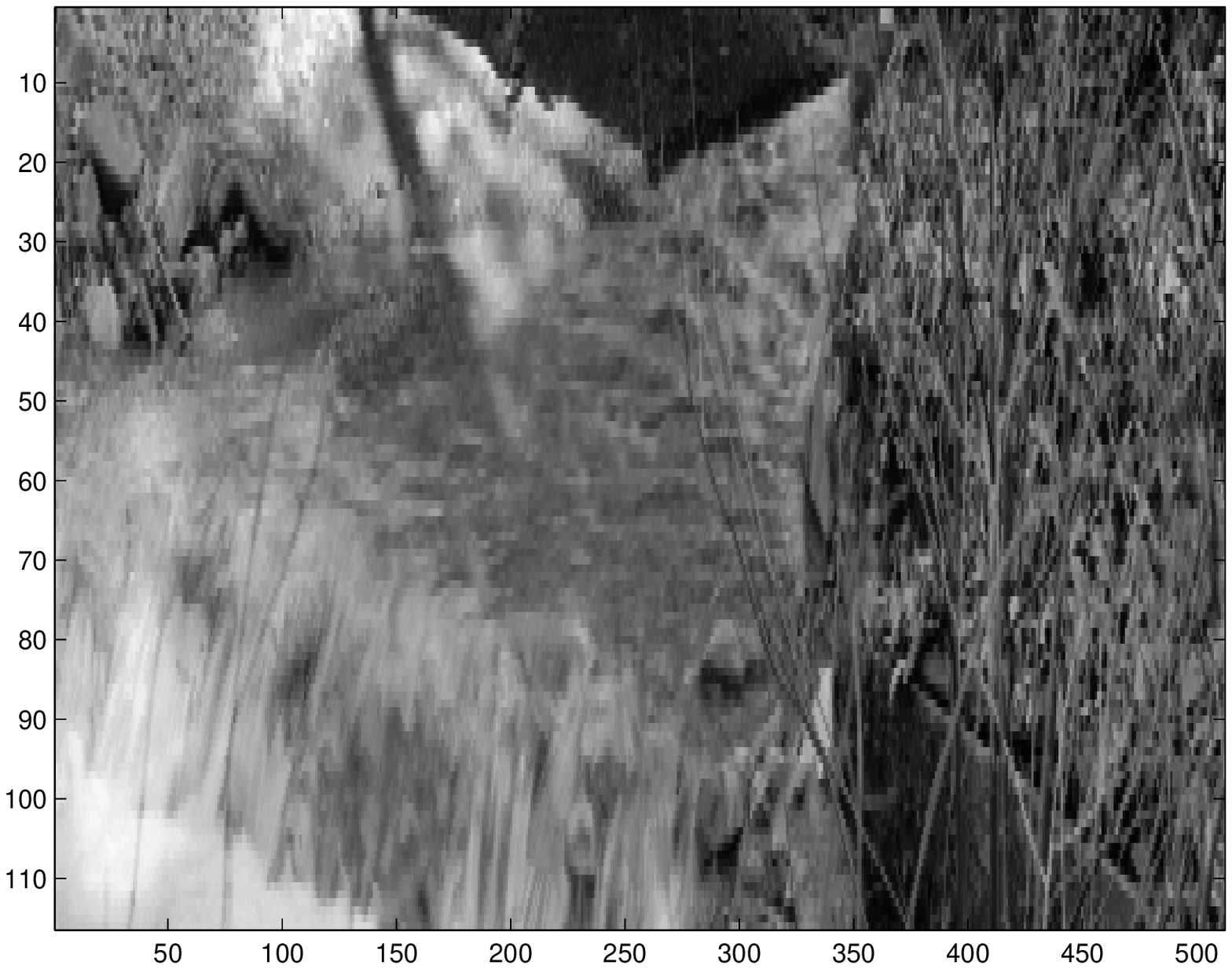} \\
  (e) {\small Signal $X^{(225)}.$} &  (f) {\small Signal $X^{(280)}.$}\\
     \includegraphics[width=8cm,height=2.5cm]{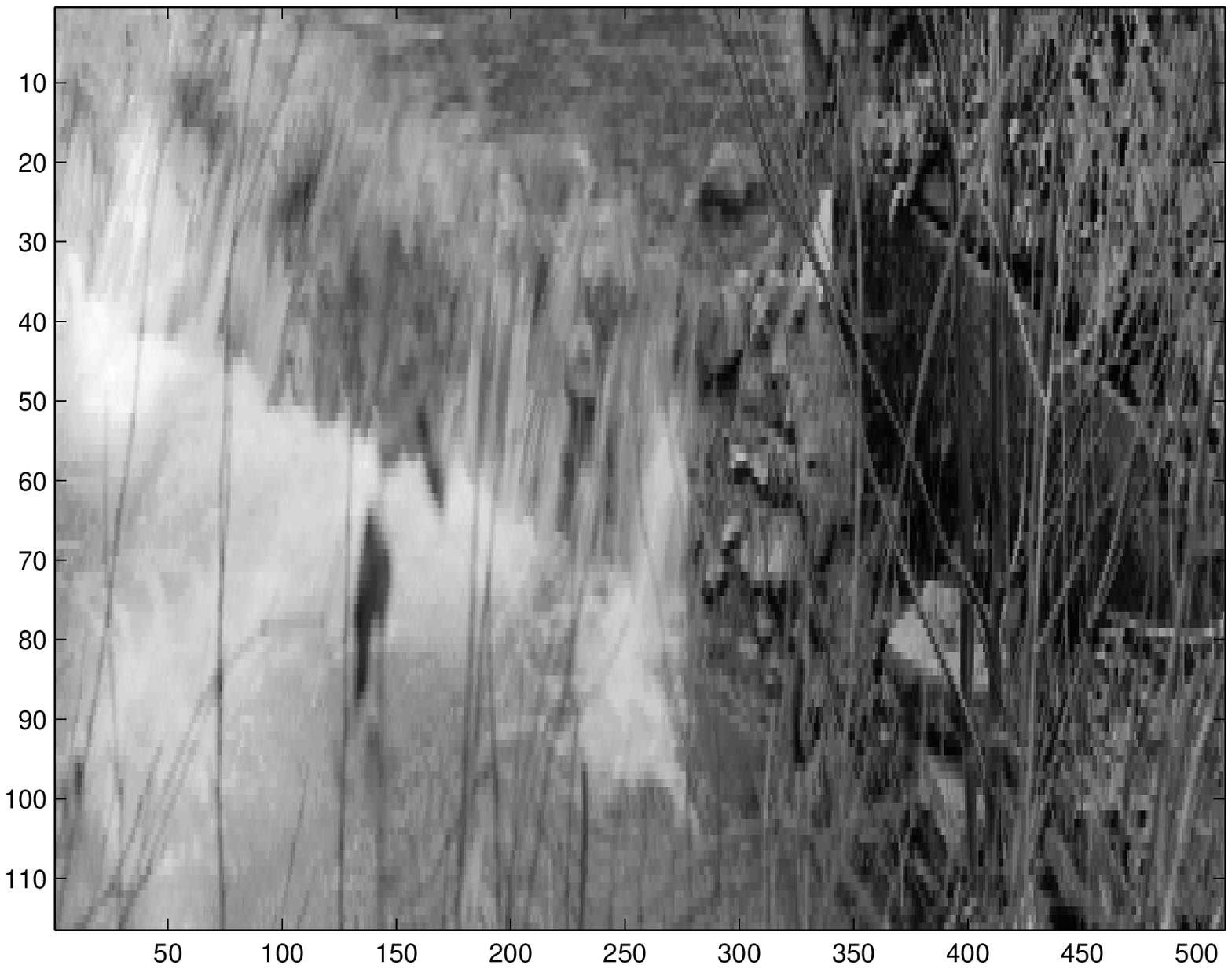} &   \includegraphics[width=8cm,height=2.5cm]{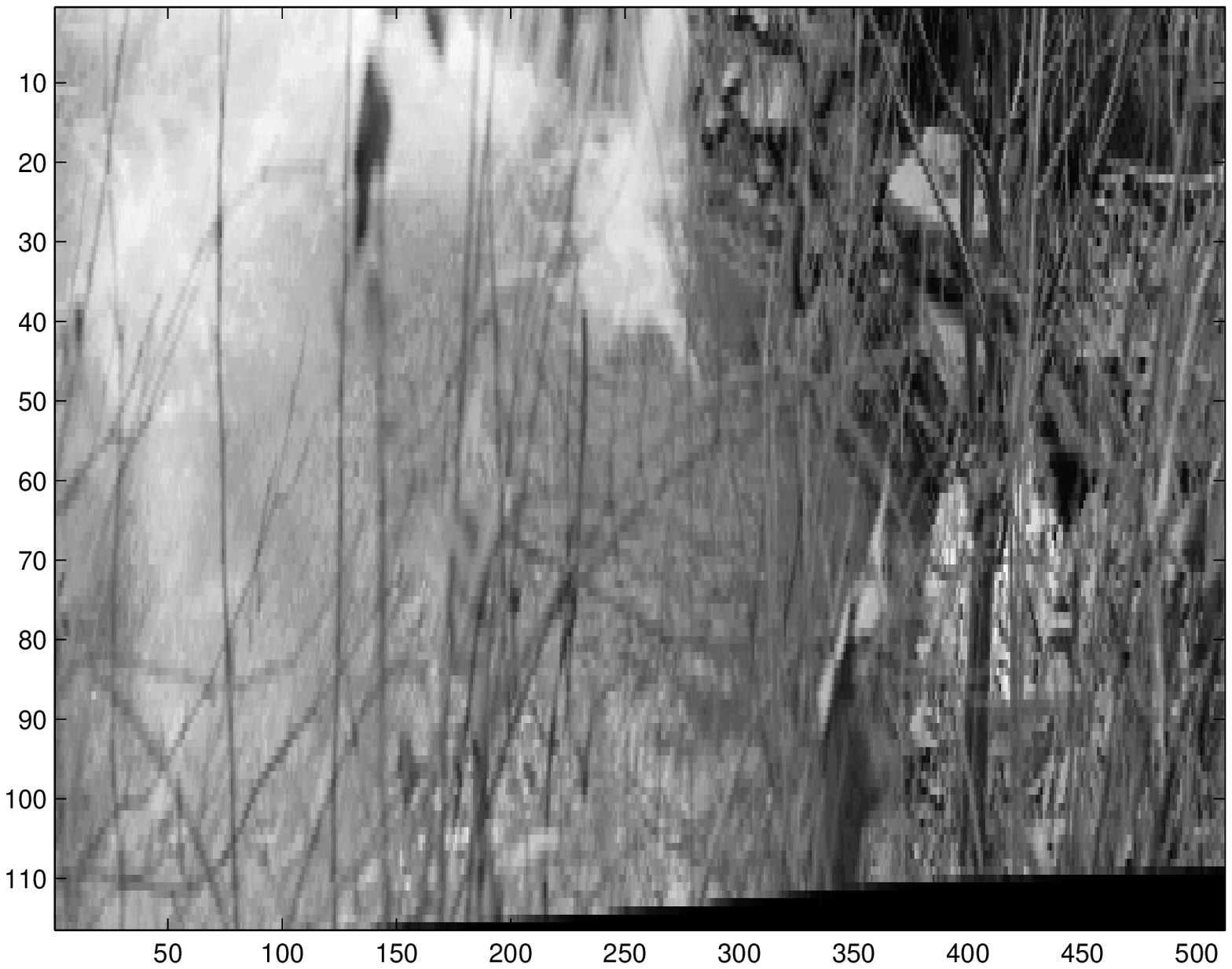} \\
  (g) {\small Signal $X^{(340)}.$} &  (h) {\small Signal $X^{(400)}.$}
\end{tabular}
 \vspace*{-5mm}\caption{Examples of selected reference signals.}
 \label{fig1}
 \end{figure}
\end{center}

\hspace*{-10mm}\begin{figure}[h]
\centering
 \vspace*{-5mm}\begin{tabular}{c@{\hspace*{1mm}}c}
    \includegraphics[width=8cm,height=2.5cm]{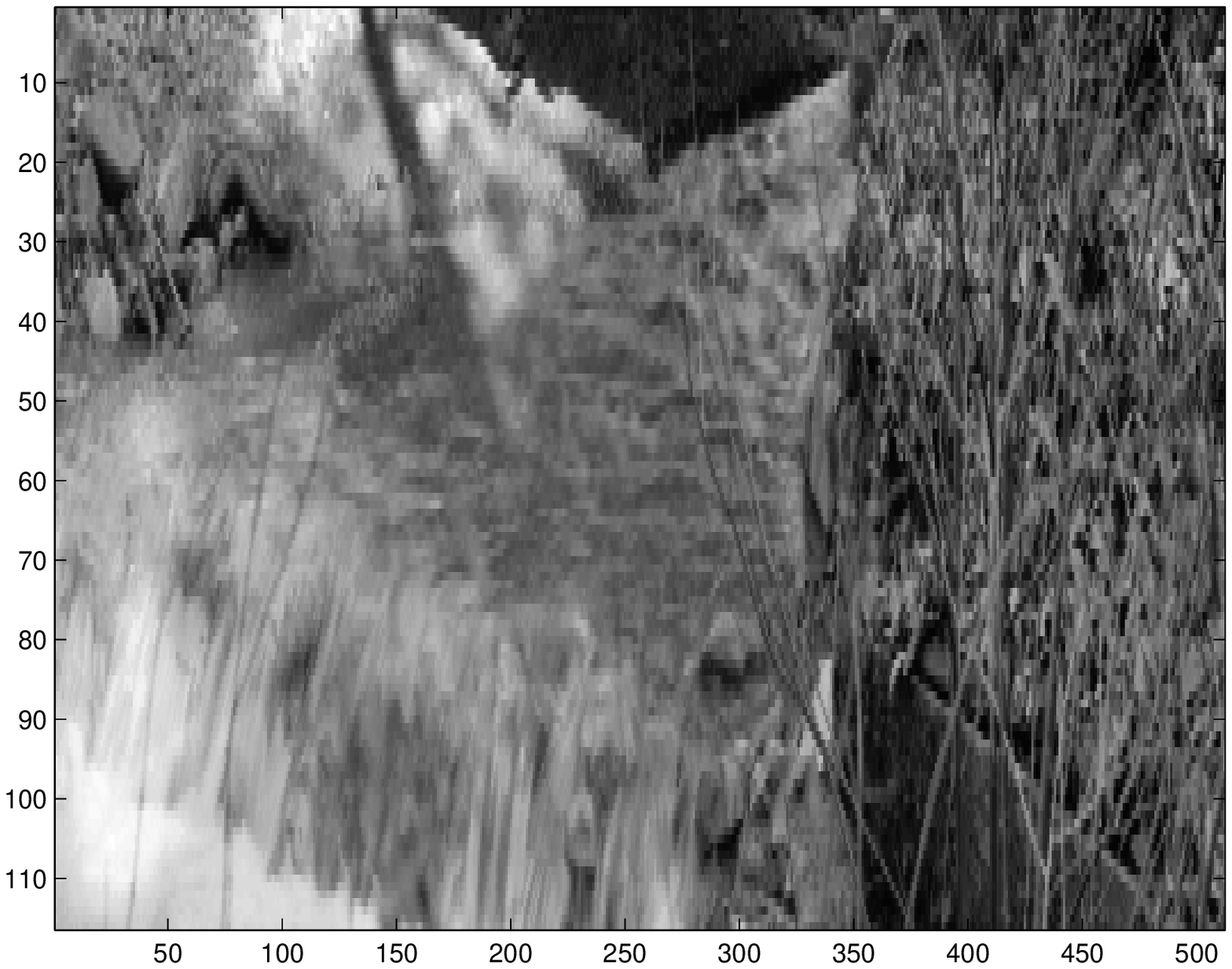} &   \includegraphics[width=8cm,height=2.5cm]{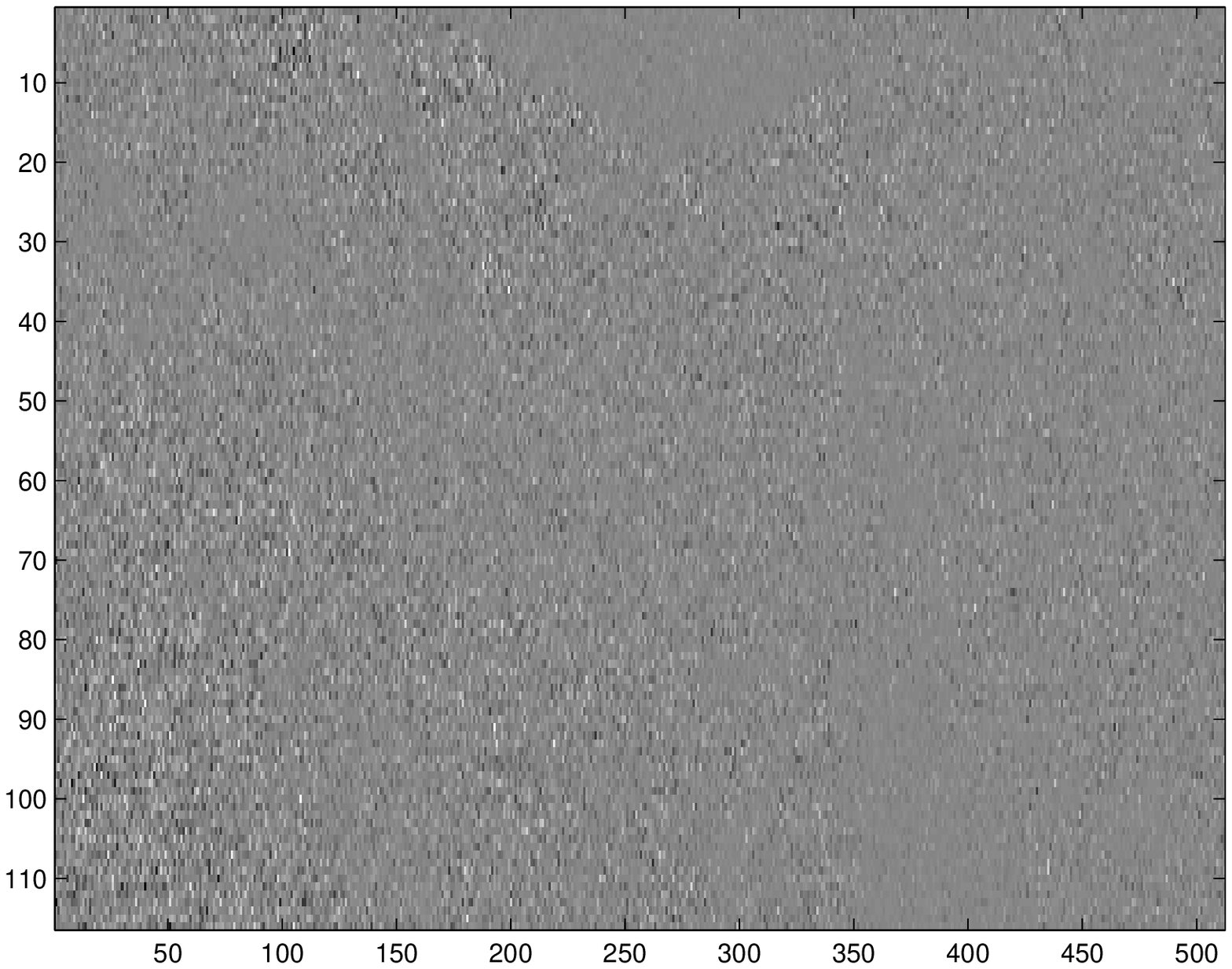} \\
 {\small (a)} & {\small (b)} \\
     \includegraphics[width=8cm,height=2.5cm]{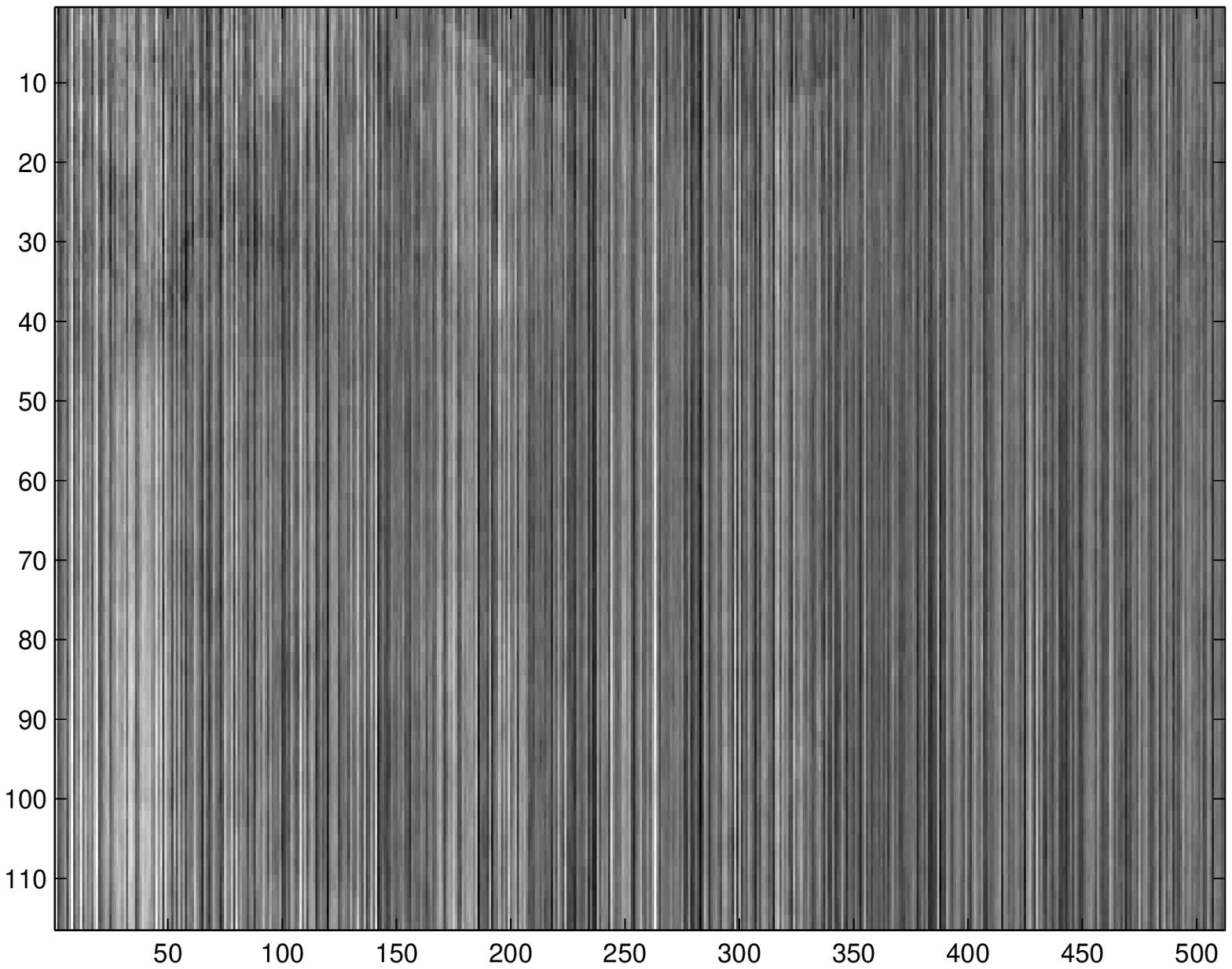} &   \includegraphics[width=8cm,height=2.5cm]{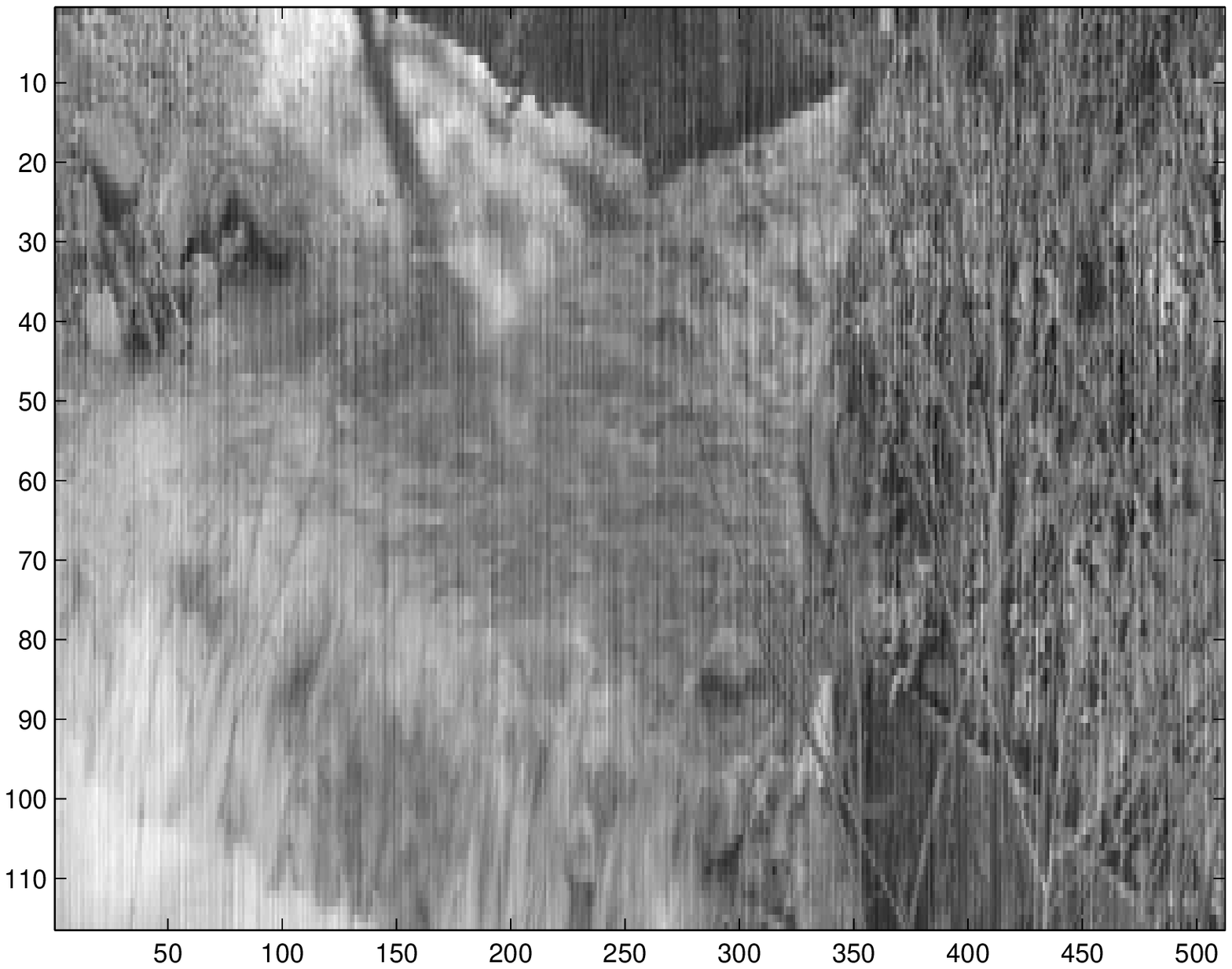} \\
 {\small (c)} & {\small (d)}
\end{tabular}
 \vspace*{0mm}\caption{Illustration of simulation results  from  Tables 1 and 2. (a): Reference signal  $X^{(280)}.$ (b): Observed data $Y^{(280)}$. (c): Reconstruction of $X^{(280)}$ from $Y^{(280)}$  by KLT with  compression ratio $10/116$. (d):  Reconstruction of $X^{(280)}$ from $Y^{(280)}$  by proposed transform $F$ with  compression ratio $10/116$.}
 \label{fig2}
 \end{figure}

\section{Simulations}\label{sim}


Here, we consider a case where $K_{_X}$ and $K_{_Y}$ (introduced in Section \ref{some}) are represented by finite signal sets with $N$
members, and illustrate the advantages of the proposed technique over  methods based on the KLT approach.
In many practical problems (arising, e.g, in a DNA analysis \cite{fri1}) the number $N$ is quite large, for instance,
$N=\mathcal O(10^4)$  \cite{fri1}. In these simulations, we set $N=400$.

Let us suppose that $\y(\tau_{1},\cdot)$, $\y(\tau_{2},\cdot)$, $\ldots$, $\y(\tau_{400},\cdot)$ are $400$  input
stochastic signals    where  $\y(\tau_{k},\cdot)\in L^2(\Omega, \rt^n)$, for $k=1,\ldots,400$ and $n=116$. Reference
stochastic signals are
$\x(\tau_{1},\cdot),$ $\x(\tau_{2},\cdot), \ldots, \x(\tau_{{400}},\cdot)$, where $\x(\tau_{k},\cdot)\in
L^2(\Omega, \rt^m)$, for $k=1,\ldots,400$ and $m=n=116$.
Thus, in these simulations, the interval $[a \hspace*{1mm} b]$ introduced in Section \ref{some} is modelled as 400
points $\tau_{k}$ with $k=1,\ldots,400$ so that $[a \hspace*{1mm} b] = [\tau_{1}, \tau_{2}, \ldots,  \tau_{{400}}]$.

Signals $\x(\tau_{k},\cdot)$ and  $\y(\tau_{k},\cdot)$ have been simulated as digital images presented by $116 \times 512$
matrices $X^{(k)}$ and $Y^{(k)}$, respectively, with $k=1,\ldots, 400$. Each column of matrices $X^{(k)}$ and $Y^{(k)}$ represents a realization of
signals $\x(\tau_{k},\cdot)$ and $\y(\tau_{k},\cdot)$, respectively.
Each matrix $X^{(k)}$ represents  data  obtained from a digital photograph `Stream and bridge'\footnote{The database
is available in http://sipi.usc.edu/services/database.html.}. Examples of selected images $X^{(k)}$ are shown in
Fig. \ref{fig1}.

Observed noisy images $Y^{(1)}, \ldots, Y^{(400)}$ have been simulated in the form
$$
Y^{(k)} =  \mbox{\tt randn}\bullet X^{(k)}\bullet \mbox{\tt rand},
$$
for each $k=1,\ldots,400$. Here,
$\bullet$ means the Hadamard product, and $\mbox{\tt randn}$ and $\mbox{\tt rand}$ are $116\times 512$ matrices
 with random entries, and they simulate noise. The entries of $\mbox{\tt randn}$ are  normally distributed with mean zero, variance
 one and standard deviation one.  The entries of $\mbox{\tt rand}$  are  uniformly distributed in the interval
 $(0, 1)$. A typical example of such noisy images is given in Fig. \ref{fig2} (b).

We wish to filter and compress the  observed data $Y^{(1)}$,  ..., $Y^{(400)}$ so that their subsequent reconstructions
would be close to the reference signals $X^{(1)}$, ..., $X^{(400)}$, respectively.

The proposed transform $\f$ given by (\ref{fyd1})--(\ref{fydj1}) and (\ref{ffg1})--(\ref{ggj1}), the generic KLT \cite{tor100} and the transform \cite{tor1} were applied to this task. To compare
their performance, we address three issues as those in Sections \ref{mot} and \ref{con}, as follows.

\subsection{Large sets of signals}\label{}

In these simulations, the signal sets are large but finite. Therefore, the known transforms developed for compression
of an {\em individual} {\em finite} random signal-vector can be applied to this case. This issue has been mentioned in Section
\ref{mot} above.
Nevertheless the known methods imply difficulties (an insufficient accuracy and excessive computational work) that
are discussed below.

\subsection{Associated accuracy and compression ratios}\label{}

\subsubsection{Proposed transform $\f$} As it has been mentioned in Section \ref{degree}, the proposed technique has two degrees of freedom, a number $p$ of
interpolation pairs $\{\x_j, \y_j\}_{j=1}^p$ and  rank $r_j$ of matrix $G_j$ (see (\ref{r1}), (\ref{b0j}) and (\ref{rr1})).
We set $r=r_j$ for all $j=1,\ldots, 400$.

To demonstrate properties and advantages of the proposed transform $\f$, interpolation pairs $\{X_1, Y_1\},\ldots, \{X_{p}, Y_p\}$ have been chosen in different
ways as follows:
\begin{description}
\item[\normalfont $1$st choice:]\hspace*{5mm} $p=9$ and interpolation pairs are $\{X_1, Y_1\}$ $=\{X^{(1)}, Y^{(1)}\}$,
$\{X_{i+1}, Y_{i+1}\}$ $=\{X^{50i}, Y^{50i}\}$ for $i=1,\ldots,8$;
\item[\normalfont $2$nd choice:]\hspace*{6mm} $p=21$ and  interpolation pairs are $\{X_1, Y_1\}$ $=\{X^{(1)}, Y^{(1)}\}$,
$\{X_{i+1}, Y_{i+1}\}$ $=\{X^{20i}, Y^{20i}\}$ for $i=1,\ldots,20$;
\item[\normalfont $3$rd choice:]\hspace*{6mm} $p=41$ and   interpolation pairs are $\{X_1, Y_1\}$ $=\{X^{(1)}, Y^{(1)}\}$,
$\{X_{i+1}, Y_{i+1}\}$ $=\{X^{10i}, Y^{10i})$ for $i=1,\ldots,40$;
\item[\normalfont $4$th choice:]\hspace*{6mm} $p=81$  and  interpolation pairs are $\{X_1, Y_1\}$ $=\{X^{(1)}, Y^{(1)}\}$,
 $\{X_{i+1}, Y_{i+1}\}$ $=\{X^{5i}, Y^{5i}\}$ for $i=1,\ldots,80$;
\end{description}
For  $k=1,\ldots,400$, the accuracy associated with compression, filtering of each $Y^{(k)}$ and its subsequent reconstruction
by the proposed transform $F$ is represented by
  \begin{equation}\label{erk1}
  \varepsilon_k (F)=\|X^{(k)} - {F}[Y^{(k)}] \|^2_F
\end{equation}
and
\begin{equation}\label{vare1}
\hspace*{-2mm}\varepsilon_{\min} (F)= \min_{k=1,\ldots,400} \varepsilon_k (F),\hspace*{1mm}
\varepsilon_{\max} (F)= \max_{k=1,\ldots,400} \varepsilon_k (F),
\end{equation}
In Fig. \ref{fig2} (d), an example of the restoration of signal $X^{(280)}$ from noisy observed data $Y^{(280)}$ is given.
$X^{(280)}$ and $Y^{(280)}$ are typical representatives of signals under consideration.
The image in Fig. \ref{fig2} (d) has been evaluated for $p=9$ as in the $1$st choice above and the compression ratio $10/116$.

Values of $\varepsilon_{\min} (F)$ and $\varepsilon_{\max} (F)$ associated with different choices of the above interpolation pairs
are given in Table 1. In the first column, the compression ratios used in the  transform $F$ are given. In particular, it follows
from Table 1 that the accuracy improves when the number of interpolation pairs increases. This is a confirmation of the statement
(\ref{ertf})--(\ref{pinf}) of Theorem \ref{thm3}.

\subsubsection{Individual KLTs $\cite{tor100}$}  To each pair $X^{(k)}$, $Y^{(k)}$, an individual KLT $K_k$
has been applied, where $k=1,\ldots,400$. Thus, the KLT $K_k$ has to be applied 400 times.
In Fig. \ref{fig2} (c), an example of the restoration of signal $X^{(280)}$ by the KLT with the compression ratio $10/116$ is given.

The error associated with compression, filtering of each $Y^{(k)}$ and its subsequent reconstruction  by the KLT  is represented by
  \begin{equation}\label{erk2}
\varepsilon  (K_k)=\|X^{(k)} - {K_k}[Y^{(k)}]\|^2_F.
\end{equation}
To compare the proposed transform $F$ and the KLT, we denote
\begin{equation}\label{varf1}
\Delta_{{\max}} = \max_{k=1,\ldots,400}[\varepsilon  (K_k)/\varepsilon_k (F)]
\end{equation}
and
\begin{equation}\label{varf2}
\Delta_{{\min}} = \min_{k=1,\ldots,400} [\varepsilon  (K_k)/\varepsilon_k (F)].
\end{equation}
In other words, $\Delta_{{\max}}$ and  $\Delta_{{\min}}$ represent maximal and minimal magnitudes of the ratios of the
accuracies associated with $F$ and the KLT $K_k$. These ratios  have been calculated with  the same two ranks of $F$ and $K_k$, $r=5$ and
$r=10$, i.e. with the same two compression ratios of $F$ and $K_k$, $c=5/116$ and $c=10/116$.

The results are presented in Table 2.
It follows from Table 2 that our transform $F$ provides the substantially better associated accuracy. Depending on the number of
interpolation pairs $p$ and compression ratio $c$, the accuracy associated with $F$ is from 4 to  25  times better than
that of the KLT.

\subsubsection{Generic KLT} \cite{tor1}. In these simulations, the generic KLT \cite{tor1} provides the associated accuracy that
is worse than that provided by the individual KLTs above.

\subsection{Computational work}\label{comp11}

The proposed transform requires the computation of $p-1$ pseudo-inverse matrices (in (\ref{b0j}), with $M_{G_j}=\oo$), $p-1$  SVDs
in (\ref{svd-ej}) and $3(p-1)$ matrix multiplications in (\ref{b0j}) and (\ref{svd-ej}). In these simulations, $p=9$, $21$, $41$,
$81$.

 The individual KLTs applied, for $k=1,\ldots,400$, to each pair $\x_k:=\x(\tau_{k},\cdot)$,  $\y_k:=\y(\tau_{k},\cdot)$ require
 the computation of 400 pseudo-inverse matrices $(E_{y_{k} y_{k}}^{1/2})^{\dag}$ and 400 SVDs of matrices
$\langle\langle E_{x_{k}y_{k}}(E_{y_{k}y_{k}}^{1/2})^{\dag}\rangle\rangle_{r_k} (E_{y_{k}y_{k}}^{1/2})^{\dag}$.
Clearly, the KLTs require substantially more computational work than that by the proposed transform.

\subsection{Summary of simulation results}\label{}

The results of the simulations confirm the theoretical results obtained above. In particular,

(i)  the accuracy $\displaystyle \er_{_F}$ associated with the proposed transform $F$ increases when
the number $p$ of interpolation pairs increases (Theorem \ref{thm3}),

(ii)  the accuracy $\displaystyle \er_{_F}$ of our transform is from 4 to 25 times better than the accuracy  associated with
the KLT-like transforms (depending on the number $p$ of interpolation pairs and the compression ratios),

(iii) the proposed transform requires less computational work than  that of the KLT-like transforms.

\section{Conclusions}

The proposed data compression technique  is constructed from {a combination} of the idea of the piece-wise linear
function interpolation  and the best rank-constrained operator approximation. This device provides the
advantages that allow us to

(i) achieve any desired accuracy in the reconstruction of compressed data (Theorem \ref{thm3}),

(ii) find  {\em a single} transform to
compress and then reconstruct any signal from the {\em infinite} signal set (Sections \ref{det1} and \ref{part}),

(iii) determine the transform in terms of
pseudo-inverse matrices so that the transform always exists (Sections \ref{det1}),

(iv) decrease the  computational load compared to
the related  techniques (Section \ref{comp11}),

(v) exploit two degrees of freedom (a number of interpolation pairs and compression ratios) to improve the
transform performance, and

(vi) use  the same  initial information (signal samples) as  is usually used in the KLT-like transforms.


\end{document}